\documentclass[preprint,pre,aps]{revtex4}
\usepackage{epsfig}
\usepackage{xcolor}

\begin{document}

\title{Structure of ionic liquids and concentrated electrolytes from a mesoscopic theory}

\author{A. Ciach}
\affiliation{Institute of Physical Chemistry, Polish Academy of Sciences, 01-224 Warszawa, Poland}
\author{O. Patsahan}
\affiliation{Institute for Condensed Matter Physics of the National Academy of Sciences of Ukraine, Lviv, Ukraine}
 \date{\today}

\begin{abstract}
Recently, underscreening in concentrated electrolytes was discovered in experiments and confirmed in simulations and theory. It was found that the correlation length of the charge-charge correlations, $\lambda_s$, satisfies the scaling relation $\lambda_s/\lambda_D\sim (a/\lambda_D)^n$, where $\lambda_D$ is the Debye screening length and $a$ is the ionic diameter. However, different values of n were found in different studies. In this work we solve this puzzle within the mesocopic theory that yielded n=3 in agreement with experiments, but only very high densities of ions were considered [A. Ciach A. and O. Patsahan, {\it J.Phys.: Condens. Matter} {\textbf 33}, 37LT01 (2021)]. Here we apply the theory to a broader range of density of ions
and find that different values of n in the above scaling can yield a fair approximation for $\lambda_s/\lambda_D$ for different ranges of $a/\lambda_D$. The experimentally found scaling holds for $2 <a/\lambda_D<4$, and we find n=3 for the same range of the reduced Debye length. For smaller  $a/\lambda_D$, we find n=2 obtained earlier in several simulation and theoretical studies, and still closer to the Kirkwood line we obtain n=1.5 that was also predicted in different works. It follows from our theory that n=3 (i.e. $\lambda_s$ is proportional to the density of ions)  when  the variance of the local charge density is large, and $\lambda_s$ is proportional to this variance times the Bjerrum length. Detailed derivation of the theory is presented. 
\end{abstract}

 \maketitle



\section{Introduction}
\label{sec: introduction}
 An interesting feature of ionic liquids and concentrated electrolytes is the hyperuniformity, meaning that the variance of the charge inside a region with the linear size $R$ grows with $R$ as $R^{d-1}$ rather than as $R^d$ for  $R\gg a$, where $d$ is the spacial dimension, and $a$ is the diameter of the ions~\cite{torquato:18:0}. These small charge fluctuations indicate significant deviations from random distribution of the ions, and follow from the charge-neutrality condition. For this reason, concentrated electrolytes and ionic liquids (IL) can be viewed as a charge-neutral background for test charges. This picture is a sort of a 'negative' of dilute electrolytes. Instead of ions screening the test charge in dilute systems,  in concentrated electrolytes we need 'holes', i.e. vacancies or solvent molecules violating the charge neutrality of the surroundings to screen the test charge. 
 The screening length $\lambda_s$ increases when the number of charge carries - ions in dilute and holes in concentrated electrolytes - decreases.  Because the density of 'holes' decreases with increasing density of ions $\rho$, the screening length in concentrated electrolytes should increase with increasing $\rho$. This intuitive, qualitative picture was confirmed by experiments~\cite{smith:16:0,lee:17:0,Groves2021}, simulations~\cite{Zeman2020}, and theory~\cite{Adar2019,ciach:21:0}.

In the experimental works \cite{smith:16:0,lee:17:0,Groves2021} it was observed that the decay length of the disjoining pressure between crossed mica cylinders confining concentrated electrolytes was  $\lambda_s/\lambda_D\sim (a/\lambda_D)^n$ with $n=3$,  where 
\begin{equation}
 \lambda_D=\sqrt{\frac{k_B T\epsilon}{4\pi e^2 \rho}} 
\end{equation}
is
the Debye screening length with $\epsilon$, $e$, $k_B$ and $T$ denoting the dielectric constant, charge, the Boltzmann constant and temperature, respectively. The same scaling relation was observed for several different systems, including alkali halide solutions, pure ionic liquids and ionic liquid  solutions. This effect has been termed as underscreening.

A similar scaling should be obeyed by the correlation length in the bulk electrolyte. Indeed, the experimentally observed scaling was predicted in Ref.~\cite{ciach:21:0} for the restricted primitive model (RPM) of hard spheres with equal diameters and equal magnitude of the charge in a structureless solvent.
 In Ref.~\cite{ciach:21:0}, the RPM was studied within
 the mesoscopic theory for inhomogeneous systems~\cite{ciach:08:1,patsahan:22:0}. However, only very high densities of ions were considered in  Ref.~\cite{ciach:21:0}, because the assumptions allowing for obtaining analytical results in this theory are valid only when the charge-density waves with the wavelength $\sim 2a$ (nearest-neighbors oppositely charged) appear with a high probability, which is the case for high densities of ions. It was shown that in such conditions, $\lambda_s$ is proportional to a variance of the charge in regions with the size $\sim a$.
   In different theories and simulations, $n=3$ was not found. Rather,  
  $n$ in the range $1<n\leq 2$
   was obtained in the scaling relation $\lambda_s/\lambda_D\sim (a/\lambda_D)^n$  \cite{Goodwin2017,Ludwig2018,Coupette2018,Adar2019,Rotenberg_2018,Coles2020,Cats2021,Zeman2020,Zeman2021,Outhwaite2021,KruckerVelasquez2021}. It is very difficult to achieve equilibrium in simulations of concentrated electrolytes at low $T$, and very long simulation runs are required. Finite size effects and noise make it difficult to observe and interpret the correlations at large distances. The results were thus obtained for not so large $\rho$ and not so low $T$.
   
 The issue of  underscreening is still under active debates. 
 Recent atomic force spectroscopy measurements for electrolyte solutions complemented  by classical density functional theory calculations for the primitive model did not demonstrate a large increase in decay length with  increasing salt concentration \cite{Kumar2022}. 
  Alternatively,  underscreening found  previously in experiments for concentrated electrolytes  \cite{smith:16:0,lee:17:0,Groves2021}  and obtained within the mesoscopic theory for the RPM \cite{ciach:21:0}  has been also found in very recent  simulations for the RPM  and supported by applying a minimal cluster theory \cite{arxiv.2209.03486}. 

  In this work we extend the studies of ref.~\cite{ciach:21:0} to a broader range of ionic densities and temperature. As the assumptions allowing for analytical results are not valid for small and medium densities of ions, here we solve our self-consistent equations for the correlation function numerically. 
 In Sec.~\ref{sec:heuristics} we consider the variance of the local charge in dilute and concentrated electrolytes. To highlight the role of the variance of the charge in regions with the size comparable with $a$ in concentrated electrolytes, we present a simplified model and estimate the correlation length for the charge correlations. In Sec.~\ref{math} we present the formalism of the mesoscopic theory applied to RPM. The results are presented in Sec.~\ref{sec:ascc} for the charge-charge correlations and in Sec.~\ref{sec:corfuro} for the density-density correlations. Our results are compared with experiment, simulations and other  theoretical results. The last section \ref{sec:conclusions} contains our  conclusions.

 \section{Variance of a local charge in dilute and concentrated electrolytes - qualitative considerations}
 \label{sec:heuristics}
\begin{figure}[!htb]
   \includegraphics[scale=0.5]{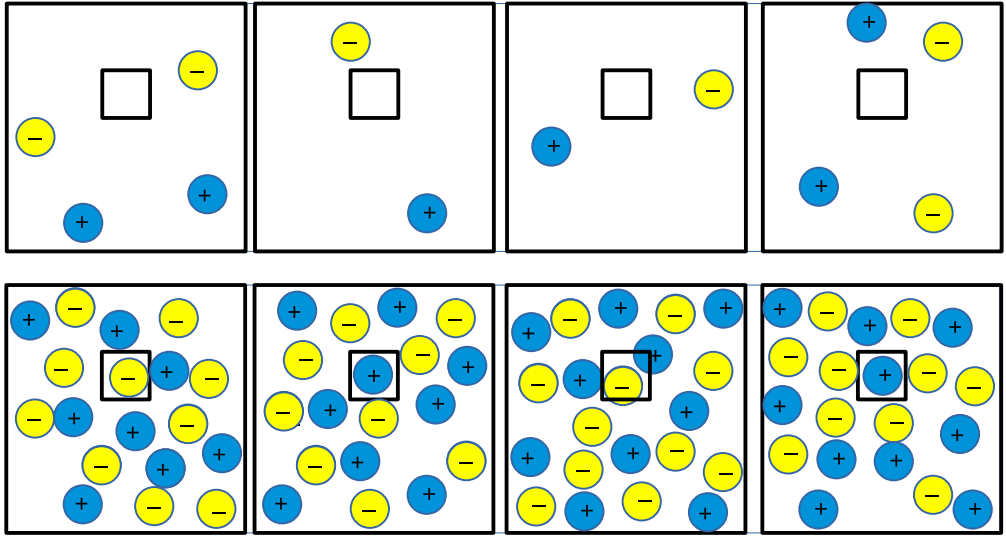}
    \caption{Cartoon showing microstates of ions in dilute (top row) and concentrated (bottom row) electrolytes. The solvent molecules are not shown for clarity. The small square inside the system represents a mesoscopic subsystem. In the dilute electrolyte the subsystem is typically uncharged, while in the concentrated one, very often either  an anion or  a cation is present inside it.
   The considered subsystem  is therefore either negatively or positively charged, even though the charge averaged over all microstates vanishes. The variance of the charge, however, in the bottom row is large. }
\label{fig:snapshots}
   \end{figure}

 In contrast to the small variance of the charge in regions with $R\gg a$, the variance of the charge in a subsystem with $R<2a$ is large  if the  electrolyte is concentrated. As illustrated in Fig.~\ref{fig:snapshots},  in dilute electrolytes any small subsystem contains typically solvent molecules and is therefore uncharged in majority of the microscopic states. However, the larger is the density of ions, the more often a subsystem with $R<2a$ is occupied either by an anion or by a cation. Occupancy of the small subsystem by a half of the anion and a half of the cation is rare. Thus, even though the average charge is zero, the variance of the local charge grows with the density of ions and in concentrated electrolytes becomes large. When the local charge density in the majority of the microscopic states is different from the average charge density, then the fluctuations of the local charge density play an important role and cannot be neglected. In particular, the properly calculated average energy can be significantly different from the energy calculated for the average (vanishing) charge density.  The variance of the local charge density is the main difference between dilute and concentrated electrolytes and ionic liquids (IL). While fluctuations of the local charge density can be neglected in the former case, they have to be taken into account in concentrated electrolytes and IL or IL solutions.

\begin{figure}[!htb]
   \includegraphics[scale=0.29]{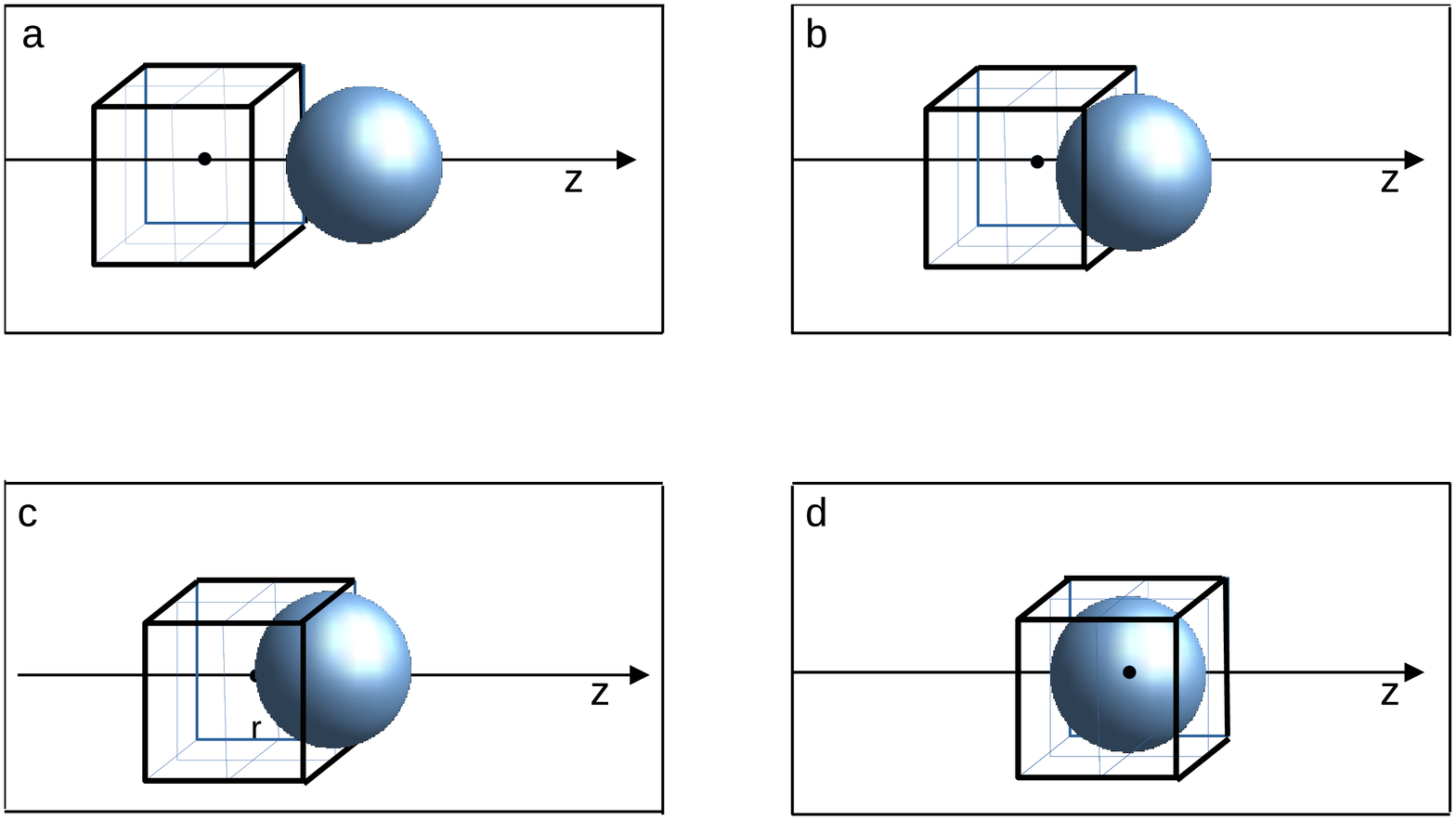}
   \includegraphics[scale=0.29]{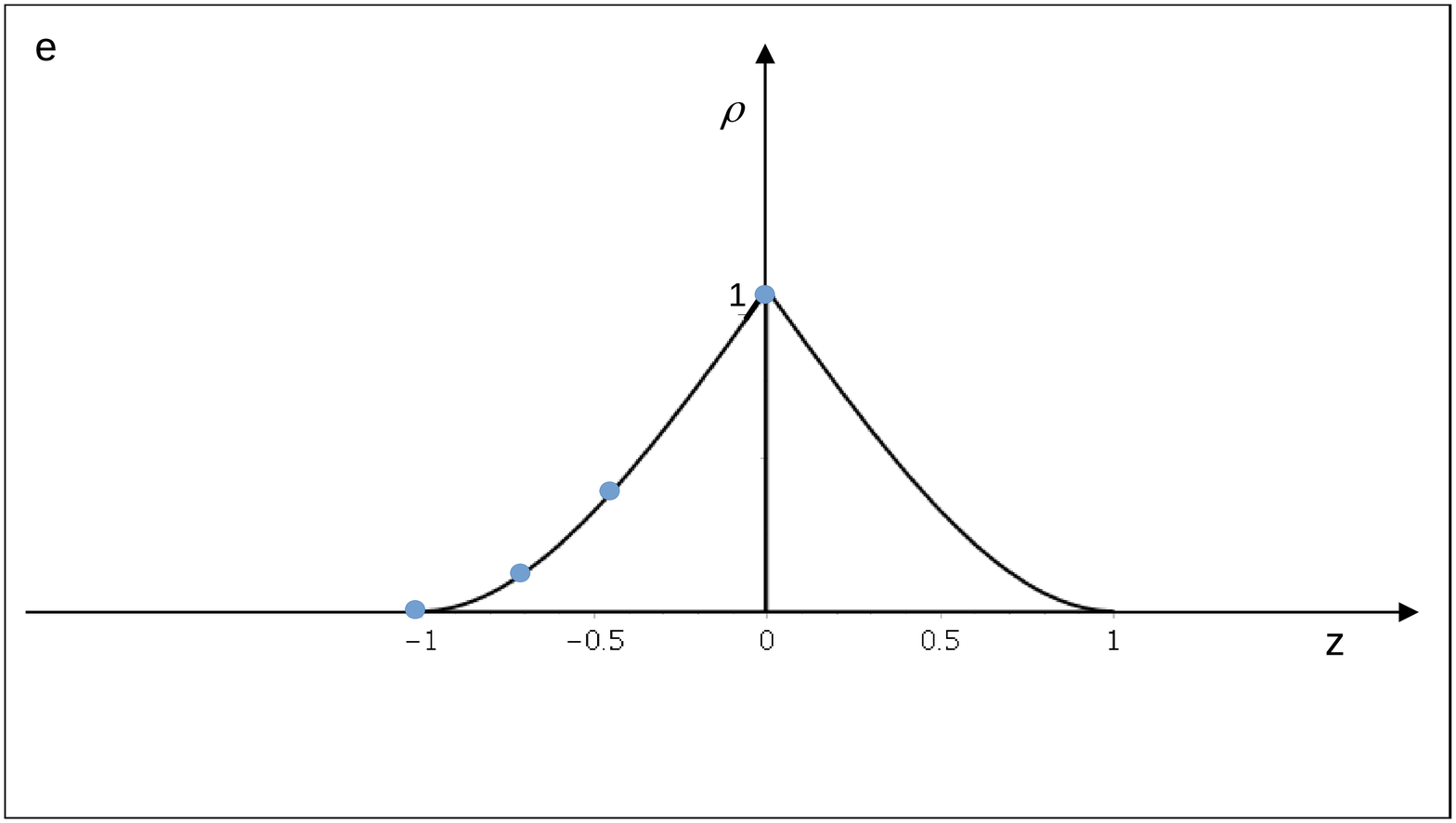}
   \caption{Illustration of the mesoscopic density for a single sphere with the diameter $a=1$ and the center at ${\bf r}=(0,0,0)$. The mesoscopic region is represented by the cube with the edge of the length 1 and the center shown as the black circle at ${\bf r}=(0,0,-1), (0,0,-0.75), (0,0,-0.5), (0,0,0)$ in panels a, b, c and d, respectively. The mesoscopic density $\rho=6\zeta/\pi$, with $\zeta$ representing the fraction of the volume of the cube that is occupied by the particles,  is shown in panel (e) as a function of $z$, with the blue circles referring to the illustrations in panels a-d.  }
\label{fig:smearing}
   \end{figure}

 In order to analyze the effect of the large variance of the local charge density on the properties of ionic systems, let us consider a local deviation $\phi({\bf r})=c({\bf r})-c$ of the charge density from the average value $c$, and its variance $\langle \phi^2\rangle$. The dimensionless charge density in the case of the same valency of the anions and the cations is given by 
\begin{equation}
\label{c}
c({\bf r})=\rho_+({\bf r})-\rho_-({\bf r}),
\end{equation}
 with $\rho_+({\bf r})$
and $\rho_-({\bf r})$ denoting the dimensionless local density of the cations and the anions.
The local dimensionless number density of the ions is defined by
 \begin{equation}
 \rho({\bf r})=\rho_+({\bf r})+\rho_-({\bf r}).
 \end{equation}
  In a macroscopic system,  $\rho_i=a^3N_i/V$, with $i=+,-$, and $N_i$ denoting the number of $i$-th type ions inside the macroscopic volume $V$. 
The local density should be defined in a somewhat different way, because the mesoscopic volume can be occupied by a fraction of an ion (see Fig.~\ref{fig:snapshots}). Thus, we define the local densities by $\rho_i({\bf r})=6\zeta_i({\bf r})/\pi$, where $\zeta_i({\bf r})$ is the fraction of the mesoscopic volume around ${\bf r}$ that is occupied by the ion of the $i$-th type. The construction of $\zeta_i$ is illustrated in Fig.~\ref{fig:smearing}.
We assume that the charge $e$ is uniformly distributed over the volume of the ion, and $ec({\bf r})$ is a continuous charge density.
In the disordered phase and in absence of any boundaries or external fields, $\langle c\rangle=0$ and  $\langle \phi^2\rangle$ is position independent. 
   
\begin{figure}[!htb]
   \includegraphics[scale=0.4]{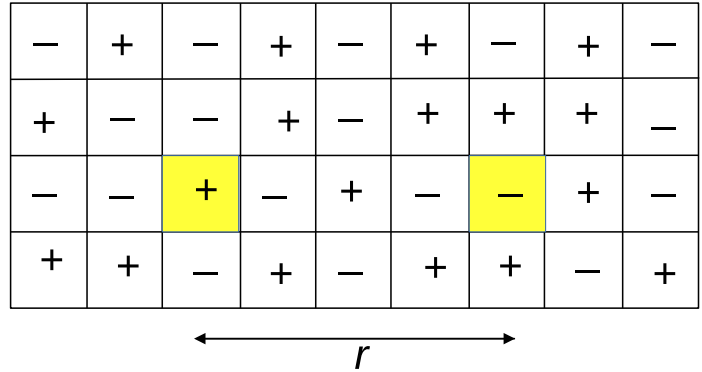}
   \caption{The concentrated electrolyte or IL divided into cells with the size $a\le R<2a$, and a representative distribution of the sign of the standard deviation of the local charge from zero, $\pm\sqrt{\langle\phi^2\rangle}$. If we require that the selected cells separated by the distance $r$ are oppositely charged, then both the energy and the entropy of the system decrease, giving negative or positive excess free energy for $r$ smaller or larger than $l_B\langle\phi^2\rangle/\ln 2$, respectively.  }
\label{fig:lattice}
   \end{figure}

Let us divide the system into cells with the linear size $a\le R<2a$.  We expect that in concentrated electrolytes the standard deviation of the charge from zero in each cell  is equal to $+\sqrt{\langle \phi^2\rangle}$ or  $-\sqrt{\langle \phi^2\rangle}$. A representative distribution of the positive and negative sign of $\sqrt{\langle \phi^2\rangle}$  among the cells  in our simplified model of a concentrated electrolyte is shown in Fig.\ref{fig:lattice}.
The distribution of the + and - sign among the cells is governed by the competition between the entropy favoring the random distribution of the signs, and the energy favoring oppositely charged neighbors. Let us focus on the two distinguished cells in Fig.~\ref{fig:lattice} separated by the distance $r^*=r/a$. We shall measure the distance  in $a$-units in the whole article, but the asterisk will be omitted for clarity of the notation. There are four possible pairs of signs in these two cells, [+,+], [-,-], [+,-] and [-,+]. If we require that the selected cells are oppositely charged, the  internal energy in $k_BT$ units decreases by $-l_B\langle \phi^2\rangle/r $
where 
\begin{equation}
    l_B = \frac{ e^2}{k_BT\epsilon a}=\frac{1}{4\pi \rho\lambda_D^2 a}
\end{equation}
 is the Bjerrum length, i.e. the distance between the ions at which the Coulomb potential equals the thermal energy $k_BT$, in $a$-units.

 The decrease of the energy is accompanied by a decrease of the entropy - two less states of the two cells are possible - and the excess free energy associated with fixing opposite charges in the cells separated by the distance $r$ is 
 \begin{equation}
 \beta\Delta F\approx \frac{-l_B\langle \phi^2\rangle}{r}+\ln 2,
 \end{equation}
 where $\beta=1/(k_BT)$.
 We can easily see that $\beta\Delta F<0$ for $r<l_B\langle \phi^2\rangle/\ln 2$, and the energy wins,
 while for  distances larger than $l_B\langle \phi^2\rangle/\ln 2$ the entropy dominates. It is thus favourable to correlate the charges up to the distance $r\propto l_B\langle \phi^2\rangle$. This distance can be considered as a rough estimation of the correlation length for the charge-charge correlations. In large systems the variance of the fluctuating quantity is  proportional to the number of the fluctuating objects, therefore we conclude that the correlation length $\lambda_S$ should be approximately proportional to $l_B\rho$. However, the estimation $ l_B\langle \phi^2\rangle\propto l_B\rho$ cannot be exact for the considered mesoscopic regions, especially when  $\rho$ is not large.

In the mean-field (MF) approximation, the average quantities such as the internal energy, are calculated on the basis of average densities of the components. 
MF works very well when in majority of the microscopic states the local densities are equal or close to the average densities. In the case of a large variance of the local densities, however, the energy in a large number of the microscopic states can be significantly different from the energy in the microstates with the densities equal or close to the average densities. In this case the averaging of the energy with the appropriate probability distribution can lead to a result quite different from the energy calculated for the average densities. For the remaining average quantities, including correlation at large distances, similar strong effect of the variance of the local densities can be expected. Thus, the fluctuation contribution to the grand thermodynamic potential should be taken into account in all the systems with large variance of the local densities.

\section{Mathematical formalism of the mesoscopic theory for electrolytes and IL}
\label{math}
In this section we summarize the formalism developed in a series of works and applied to different systems with competing interactions \cite{ciach:08:1,ciach:11:2,ciach:18:0,patsahan:22:0}. We consider the mesoscopic densities $\rho_i$ discussed in the previous section that for the ionic systems are more convenient than the local volume fractions, but in fact differ from the latter only by the factor $6/\pi$. A particular form of $\rho_i({\bf r})$  can be induced by external fields $h_i({\bf r})$, and the grand potential functional of $h_i$ can be written in the form  
\begin{equation}
\label{Omega}
 \beta\Omega_h[\{h_i\}]=-\ln \int D\rho_+\int D\rho_- 
 \exp\Big(-\beta\Omega_{co}[\{\rho_i\}]+\beta\int d{\bf r}h_i({\bf r})\rho_i({\bf r})\Big).
\end{equation}
 The summation convention for repeated indexes is used here and below.
 In (\ref{Omega}),
 the integration over the mesoscopic degrees of freedom, $\rho_i$, 
and the integration over the microscopic states for each fixed  $\rho_i$ are performed separately.  $\exp(-\beta\Omega_{co}[\{\rho_i\}])$ is equal to 
  $\exp(-\beta H)$ integrated  over all the microscopic states compatible with given $\rho_i$, with $H$ denoting the microscopic Hamiltonian. Because $\Omega_{co}[\{\rho_i\}]$ is calculated for fixed $\rho_i$, i.e. with suppressed mesoscopic fluctuations, we can use the MF approximation for it. 
  According to the general thermodynamic formula, 
    \begin{equation}
    \label{Omco}
    \Omega_{co}[\{\rho_i\}]=
   U_{co}[\{\rho_i\}]-TS_{co}[\{\rho_i\}] -\mu\int d{\bf r}\rho({\bf r}) ,
   \end{equation}
where $ U_{co}[\{\rho_i\}]$  and $S_{co}[\{\rho_i\}]$ are the internal energy 
and the entropy, respectively, in the presence of the constraints $\{\rho_i\}$  imposed on the microscopic states, and $\mu$ is the chemical potential of the ions.  We assume 
$-TS_{co}=\int d{\bf r} f_h(\rho_+({\bf r}),\rho_-({\bf r}))$, where
\begin{equation}
\label{fh}
\beta f_h=\rho_+\ln \rho_++\rho_-\ln\rho_-+\beta f_{ex}(\rho)
\end{equation}
 is the free-energy per unit volume 
of the hard-core reference system in the local-density approximation. The first two terms come from the entropy of mixing, and the last term describes packing of hard cores. 
For the internal energy we postulate 
\begin{equation}
    \label{Uco}
    U_{co}[\{\rho_i\}]=
    \frac{1}{2}\int d{\bf r}_1\int d{\bf r}\rho_i({\bf r}_1)V_{ij}(r) g_{ij}(r)\rho_j({\bf r}_1+{\bf r}),
\end{equation}
where  $r=|{\bf r}|$ and $V_{ij}(r)$ is the interaction between the ions of the $i$-th and $j$-th type, and    $g_{ij}(r)$ is the pair distribution function.
In general, the interaction potential consists of the Coulomb potential and possible additional interactions. In the case of suppressed mesoscopic fluctuations, we assume $g_{ij}(r)=\theta(r-1)$, where $\theta$ is the unit step function, and $r$ is in $a$-units. With this assumption, we avoid contributions to the internal energy from overlapping hard cores of the ions, and have $g_{ij}(r)=1$ for $r\to \infty$. If only the Coulomb potential is taken into account, then we obtain the simple formula
\begin{equation}
\label{Ucoc}
\beta U_{co}=\frac{l_B}{2}\int d{\bf r}_1\int d{\bf r} c({\bf r}_1)\frac{\theta(r-1)}{r}c({\bf r}_1+{\bf r})=\frac{l_B}{2}\int \frac{d{\bf k}}{(2\pi)^3} \frac{4\pi \cos k}{k^2}  \hat c({\bf k})\hat c(-{\bf k}) ,
\end{equation}
 where  $\hat c({\bf k})$ denotes the function $c$ in Fourier representation,  and $k=|{\bf k}|$.
 We will use the same convention (a hat) for all functions in Fourier representation in the 3 dimensional space. 
  
  Once the form of $\Omega_{co}$ is assumed, we can return to the generating functional of the correlation functions, (\ref{Omega}). 
  The average mesoscopic density at ${\bf r}$, and the correlation function between fluctuations of $\rho_i$ in the mesoscopic regions around ${\bf r}_1$ and ${\bf r}_2$  are given by
\begin{equation}
 \bar \rho_i({\bf r})=\frac{\delta(-\beta \Omega_h)}{\delta(\beta h_i({\bf r}))}
\end{equation}
and
\begin{equation}
\label{Gij}
G_{ij}({\bf r}_1-{\bf r}_2)=\langle\rho_i({\bf r}_1)\rho_j({\bf r}_2)\rangle-\bar \rho_i({\bf r}_1)\bar \rho_j({\bf r}_2)=\frac{\delta \bar \rho_i({\bf r}_1)}{\delta \beta h_j({\bf r}_2)}.
\end{equation}

   The Legendre transform 
\begin{equation}
\label{LT}
  \beta \Omega[\{\bar\rho_i\}]= \beta\Omega_h[\{h_i\}]+\beta\int d{\bf r}h_i({\bf r})\bar\rho_i({\bf r})
\end{equation}
  is a functional of $\bar \rho_i$, generating the inverse correlation functions,
 \begin{equation}
 \label{Cij}
 C_{ij}({\bf r}_1-{\bf r}_2)=\frac{\delta^2\beta\Omega}{\delta\bar \rho_i({\bf r}_1)\delta\bar \rho_j({\bf r}_2)}=
 \frac{\delta\beta h_i({\bf r}_1)}{\delta \bar \rho_j({\bf r}_2)}.
\end{equation}  
From (\ref{Gij}) and  (\ref{Cij}) one can easily get the analog of the Ornstein-Zernike (OZ) equation, namely ${\bf G}={\bf C}^{-1} $in the matrix sense.

 Using (\ref{Omega}), we write the functional (\ref{LT}) in the form 
 \begin{equation}
  \label{F}
   \beta \Omega[\{\bar\rho_i\}]=\beta\Omega_{co}[\{\bar\rho_i\}]-\ln \int D\phi_+\int D\phi_- \exp\Big(-\beta H_{f}[\{\bar\rho_i,\phi_i\}]\Big)
  \end{equation}
where $\phi_i({\bf r})=\rho_i({\bf r})-\bar \rho_i$ is the local fluctuation of $\rho_i$, and
\begin{equation}
\label{Hf}
  \beta H_{f}[\{\bar\rho_i,\phi_i\}]=\beta\Omega_{co}[\{\bar \rho_i+\phi_i\}]-\beta\Omega_{co}[\{\bar \rho_i\}]
  -\beta\int d{\bf r}h_i({\bf r})\phi_i({\bf r}).
 \end{equation}
  The probability that the local fluctuations $\phi_i$ appear is proportional to  $\exp(-\beta H_f)$, and the correlation functions  can be obtained from the formula
 \begin{equation}
\label{Gij2}
 G_{ij}({\bf r}_1,{\bf r}_2)=\langle\phi_i({\bf r}_1)\phi_j({\bf r}_2)\rangle=
 \frac{\int D\phi_+\int D\phi_- e^{-\beta H_{f}}\phi_i({\bf r}_i)\phi_j({\bf r}_2)}{\int D\phi_+\int D\phi_- e^{-\beta H_{f}}}.
\end{equation}
It is difficult to calculate the correlation functions from (\ref{Gij}) or (\ref{Gij2}), unless we make some approximations. 
In order to develop such an approximate theory, we focus on $C_{ij}$. 
As follows from (\ref{F}) and (\ref{Cij}), the matrix ${\bf C}$ contains a contribution from $\beta\Omega_{co}$ associated with microscopic fluctuations (the first term in (\ref{F})), and the contribution associated with the mesoscopic fluctuations  $\phi_i({\bf r})$ (the second term in (\ref{F})), and is given by the following expression
 \begin{eqnarray}
 \label{Cab}
 C_{ij}(r)=
 \frac{\delta^2 (\beta \Omega_{co})}{\delta \bar\rho_i({\bf r}_1)\delta \bar\rho_j({\bf r}_1+{\bf r})}+
 \Big\langle \frac{\delta^2 (\beta H_f)}{\delta \bar\rho_i({\bf r}_1)\delta \bar\rho_j({\bf r}_1+{\bf r})}-
 \frac{\delta(\beta H_f)}{\delta \bar\rho_i({\bf r}_1)}\frac{\delta(\beta H_f)}{\delta \bar\rho_j({\bf r}_1+{\bf r})}\Big\rangle
\end{eqnarray}
In this equation, the inverse correlation function is expressed in terms of the average of a function of $\phi_+,\phi_-$ that can be expanded in a series of correlations of different orders. 

 In the fully symmetrical case, it is convenient to calculate the charge-charge and the density-density correlations,  $G_{cc}(r)=\langle c({\bf r}_1)c({\bf r}_1+{\bf r})\rangle$ and $G_{\rho\rho}(r)= \langle \rho({\bf r}_1)\rho({\bf r}_1+{\bf r})\rangle-\bar\rho^2$,  because  $C_{c\rho}=0$ for $\bar c=0$ (vanishing external fields). In addition, from the considerations in the previous section it follows that we should take into account the variance of the local charge density.
 We shall limit ourselves to two-point correlation functions in (\ref{Cab}), and make the self-consistent Gaussian approximation
  \begin{equation}  
   \label{HG}
  \beta H_{f}[\{\bar\rho_i,\phi_i\}]\approx \beta H_G[\bar c,\bar\rho,\phi,\psi]
    \end{equation}
  with
  $\phi=c-\bar c,\psi=\rho-\bar\rho$ and
  \begin{equation} 
  \label{HG1}
  \beta H_G[\bar c,\bar\rho,\phi,\psi]=\frac{1}{2}\int d{\bf r}_1\int d{\bf r}_2
\big(  \phi({\bf r}_1)C_{cc}(r) \phi({\bf r}_2)+ \psi({\bf r}_1)C_{\rho\rho}(r) \psi({\bf r}_2)\big),
  \end{equation}
 where $C_{\alpha\beta}$ with $\alpha,\beta=c,\rho$ is a functional of $\bar c,\bar\rho$  satisfying an equation analogous to Eq.(\ref{Cab}). In principle, the term proportional to $C_{c\rho}$ should be included in (\ref{HG1}). However, we neglect this term along with the other neglected terms in $H_f-H_G$, since for $\bar c=0$ this term vanishes. We stress that $H_G$ is not equal to the Taylor expansion of $H_f$ truncated at the second order term if the fluctuation contribution in (\ref{Cab}) is present. 
 
 In the Gaussian approximation (\ref{HG})-(\ref{HG1}), we have
 \begin{eqnarray}
 \label{Ccc}
 C_{cc}(r)&=&\frac{\delta^2\beta\Omega}{\delta c({\bf r}_1)\delta c({\bf r}_1+{\bf r})}
 \\
 \nonumber
 &=&
 \frac{\delta^2 (\beta \Omega_{co})}{\delta c({\bf r}_1)\delta c({\bf r}_1+{\bf r})}+
 \Big\langle \frac{\delta^2 (\beta H_G)}{\delta c({\bf r}_1)\delta c({\bf r}_1+{\bf r})}-
 \frac{\delta(\beta H_G)}{\delta c({\bf r}_1)}\frac{\delta(\beta H_G)}{\delta c({\bf r}_1+{\bf r})}\Big\rangle
\end{eqnarray}
 with analogous equation for  $C_{\rho\rho}$.  In order to solve (\ref{Ccc}) and (\ref{HG1}) self-consistently,
  we  assume that the fluctuation contribution to the two-point functions should be taken into account according to Eq.(\ref{Ccc}), 
but for the higher-order functional derivatives of $\beta\Omega$ (i.e. the functional derivatives of $C_{\alpha\alpha}$), the
fluctuation contribution in (\ref{F}) and (\ref{Ccc}) can be disregarded. This approximation for $C_{\alpha\alpha}$ corresponds to the self-consistent one-loop approximation in the field-theoretic approach.  Based on this assumption, we make  for $n+m>2$ the approximation
\begin{equation} 
 \frac{\delta^{n+m}(\beta \Omega)}{\delta c({\bf r}_1)...\delta c({\bf r}_n)\delta \rho({\bf r}_{n+1})...\delta |rho({\bf r}_{n+m})}\approx
 \frac{\delta^{n+m}(\beta \Omega_{co})}{\delta c({\bf r}_1)...\delta c({\bf r}_n)\delta \rho({\bf r}_{n+1})...\delta \rho({\bf r}_{n+m})}.
\end{equation}
 In the local density approximation for $\beta f_h(\{\rho_i\})$, we have for $n+m>2$
\begin{equation} 
\label{dOco}
 \frac{\delta^{n+m}(\beta \Omega_{co})}{\delta c({\bf r}_1)...\delta c({\bf r}_n)\delta \rho({\bf r}_{n+1})...\delta \rho({\bf r}_{n+m})}
 =A_{m,n}(c,\rho)\delta({\bf r}_1-{\bf r}_2)...\delta({\bf r}_{n+m-1}-{\bf r}_{m+n})
\end{equation}
where
\begin{equation}
\label{Amn}
 A_{m,n}(c,\rho)=\frac{\partial^{n+m}(\beta f_h)}{\partial^n c\partial^m \rho}.
\end{equation}

It is convenient to consider the correlation functions in Fourier representation, 
because  $C_{c\rho}=0$ for $\bar c=0$, and  for $G_{cc}$ and $ G_{\rho\rho}$ defined by an equation analogous to  (\ref{Gij2}) we have $\hat G_{cc}(k)=1/\hat C_{cc}(k)$ and  $\hat G_{\rho\rho}(k)=1/\hat C_{\rho\rho}(k)$. Note that  the product of the Coulomb potential and $\theta(r-1)$ in Fourier representation is $4\pi \cos k/k^2$ that takes a negative minimum for $k_0\approx 2.46$. Thus,  charge waves with the wavelength $2\pi/k_0$ are energetically favored. This means that neighboring regions with the charge larger and smaller from zero occur with a high probability, and the variance of the local charge,
 \begin{equation}
 \label{variance}
 \langle\phi^2\rangle=\int \frac{d{\bf k}}{(2\pi)^{3}}\hat C_{cc}^{-1}(k)
 \end{equation}
should be taken into account, as already argued in sec.\ref{sec:heuristics}.
Local fluctuations of $\rho$ are not energetically favored (see (\ref{Uco}) and Fig.\ref{fig:snapshots}), large local density fluctuation are not expected. Thus,  $\langle\psi^2\rangle$ can be neglected. With the above assumptions, the equations for the inverse correlation functions in the self-consistent Gaussian approximation take the forms 

\begin{equation}
\label{Ccc3}
 \hat C_{cc}(k)\approx  \frac{4\pi l_B\cos k}{k^2} +
 A_{0,2}+\frac{A_{0,4}}{2}\langle\phi^2\rangle
\end{equation}
with  $\langle\phi^2\rangle$ given in (\ref{variance}), and
\begin{equation}  
\label{Crr3}
 \hat C_{\rho\rho}(k)\approx
 A_{2,0}+\frac{A_{2,2}}{2}\langle\phi^2\rangle  +\beta \hat V_{fl}(k),
\end{equation}
where
\begin{equation}
\label{Vfl}
\beta\hat V_{fl}(k)=-\frac{A_{1,2}^2}{2}\int d{\bf r}e^{i{\bf k}\cdot{\bf r}}G_{cc}(r)^2-\frac{A_{3,0}^2}{2}\int d{\bf r}e^{i{\bf k}\cdot{\bf r}}G_{\rho\rho}(r)^2.
\end{equation}
We took into account that in the Gaussian approximation, $\langle\phi^2({\bf r}_1)\phi^2({\bf r}_2)\rangle=2\langle\phi({\bf r}_1)\phi({\bf r}_2)\rangle^2$.
Note that $A_{m,n}$ depends only on the entropy of mixing for $n>0$, therefore $ \hat C_{cc}(k)$ does not depend on our approximation for the free energy density associated with the packing of ions. On the other hand, $\hat C_{\rho\rho}(k)$ depends on this approximation through $A_{2,0}$ and $A_{3,0}$, therefore we expect that  $\hat G_{\rho\rho}(k)$ depends on the geometry of ions more strongly than the charge-charge correlations. 

Self-consistent solutions of (\ref{variance})-(\ref{Vfl}) together with $ \hat C_{cc}(k)=1/\hat G_{cc}(k)$ and $ \hat C_{\rho\rho}(k)=1/\hat G_{\rho\rho}(k)$ give us the correlation functions. 
 We should note, however that the  integral in Eq.(\ref{variance}) must be cutoff regularized.
 The accuracy in this theory is limited by the cutoff dependence of  $\langle\phi^2\rangle$, but according to the construction of the mesoscopic theory, the cutoff $\Lambda$ is not a free parameter. The wavelengths of the charge density are limited from below because the hard cores cannot overlap, and the nearest neighbors should be oppositely charged in the charge-waves that occur with sufficiently high-probability.  We assume that consistent with the scale of coarse-graining set by the diameter of the ion, the cutoff in (\ref{variance})  should belong to a rather narrow range of values around  $\Lambda\approx 2\pi/2=\pi$. 
 As we show in the next section, the dependence on $\Lambda$ decreases with increasing $l_B\rho$.

\section{Properties of the correlation functions}
\label{sec:corfu}
\subsection{Asymptotic decay of charge-charge correlations}
\label{sec:ascc}
In order to determine the charge-charge correlations for different parts of the phase diagram, we use (\ref{Amn}) for $A_{m,n}$ and write (\ref{Ccc3}) in the form
\begin{equation}
\label{Cccs}
 \hat C_{cc}(k)\approx  \frac{4\pi l_B\cos k}{k^2} +
\frac{1}{\rho_R},
\end{equation}
where we introduced the "renormalized density" $\rho_R$ that satisfies the self-consistent equation (see (\ref{Amn})-(\ref{Ccc3}))
\begin{equation}
\label{roR}
\frac{1}{\rho_R}=\frac{1}{\rho}+ \frac{1}{\rho^3}\int_0^{\Lambda} \frac{dk}{(2\pi)^2} \frac{2k^2}{4\pi l_B\cos k /k^2+1/\rho_R}.
\end{equation}
We assume $\Lambda=\pi$ and solve (\ref{roR}) numerically.
The charge-charge correlations for $\hat C_{cc}(k)$ given by (\ref{Cccs}) were calculated in Ref.~\cite{ciach:03:1}, but with  $\rho_R$ in (\ref{Cccs}) simply equal to $\rho$, i.e. with neglected fluctuations. We can use the results of Ref.~\cite{ciach:03:1} if we replace $\rho$  by $\rho_R$ obtained from the solution of (\ref{roR}).

According to the standard pole analysis, we have
\begin{equation}
\label{Gccrp}
G_{cc}(r)=\frac{1}{2\pi r}\sum_n \frac{e^{iq_nr}q_n}{\hat C^{'}_{cc}(q_n)},
\end{equation}
where $q_n$ are simple poles with positive imaginary parts of $\hat G_{cc}$ extended to the complex $q$-plane, and $\hat C^{'}_{cc}(q_n)$ is the derivative at $q=q_n$. The poles of $\hat G_{cc}(q)$ satisfy the equation

\begin{equation}
\label{poles}
\frac{4\pi \cos q}{q^2}+S^R=0
\end{equation}
where $S^R=1/(\rho_R l_B)$, and the relevant poles are $q=i\alpha_0\pm\alpha_1$, where $1/\alpha_0$ is equal to the correlation length $\lambda_s$, and $\alpha_1$ is the wave number of the oscillations. For $S^R>S_K^R\approx 11.8$, there are two imaginary poles that satisfy the equation
\begin{equation}
\label{polesim}
\frac{4\pi \cosh \alpha_0}{\alpha_0^2}=S^R,
\end{equation}
and the monotonic asymptotic decay of correlations is determined by the smaller solution of (\ref{polesim}).
The two imaginary poles coalesce at the Kirkwood line $S^R=S_K^R\approx 11.8$, and the inverse decay length at this line satisfies $\tanh \alpha_K=2/\alpha_K$. For $S^R<S^R_K$ there is a pair of complex poles $q_{1,2}=i\alpha_0\pm\alpha_1$, and 
\begin{equation}
\label{Gccr}
G_{cc}(r)=A_{c}e^{-\alpha_0 r}\sin(\alpha_1r+\theta)/r.
\end{equation}
We should note that in this mesoscopic theory, the microscopic structure is smeared out (see Fig.~\ref{fig:smearing}), and we can only predict the decay of correlations at large distances.

\begin{figure}[!htb]
 \includegraphics[scale=0.4]{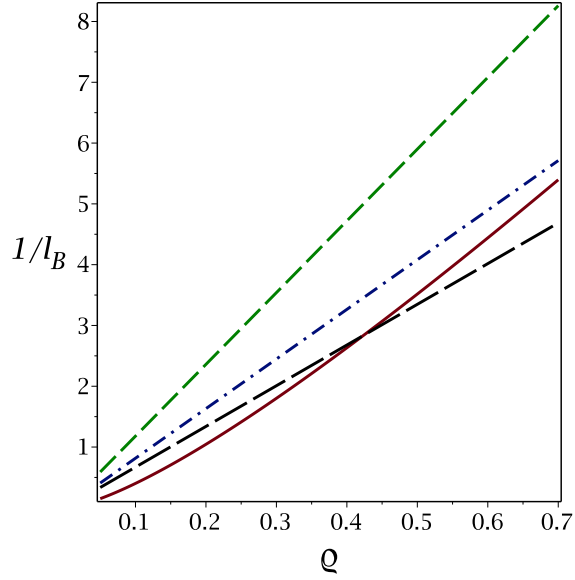}
\caption{Kirkwood line  separating the monotonic and the oscillatory decay of charge-charge correlations with the fluctuation contribution to $\hat C_{cc}$ neglected (green dashed line) and included (red solid line). Blue dash-dotted line shows the MSA result ($a/\lambda_D\approx 1.24$)~\cite{outhwaite:75:0,ciach:03:1} and  black long-dash line shows simulation result ($a/\lambda_D\approx 1.37$)~\cite{Cats2021}. $\rho$ is the dimensionless density of ions, and the Bjerrum length $l_B$ in units of the ion diameter $a$ is inversely proportional to absolute temperature.}
\label{fig:kl}
\end{figure}

We determine the Kirkwood line by a numerical solution of the equation (see (\ref{roR}))
\begin{equation}
\label{KirR}
S^R=\frac{1}{\rho l_B}+ \frac{1}{(\rho l_B)^2\rho}\int_0^{\pi} \frac{dk}{(2\pi)^2} \frac{2k^2}{4\pi \cos k /k^2+S^R}
\end{equation}
 for $S^R=11.8$.  In Fig.~\ref{fig:kl},
 $1/l_B$ at the Kirkwood line is shown as a function of $\rho$ together with our MF and mean spherical approximation results as well as with simulation results, $a/\lambda_D\approx 1.37$, obtained in Ref.~\cite{Cats2021}. Although we only expect a semi-quantitative accuracy of the cutoff-dependent result, the agreement with the simulations results is rather good for $\Lambda=\pi$,
 but our result deviates a little from a straight line. 

When the fluctuation contribution to  $\hat C_{cc}(k)$  is neglected, then  $\hat G_{cc}(k)$ diverges for $k=k_0\approx 2.46$ at the $\lambda$-line given by $(\rho_R l_B)^{-1}=|4\pi \cos k_0 /k_0^2|\approx 1.61$.  However, when the fluctuation contribution is included, the resulting equation (see (\ref{KirR}))
\begin{equation}
\label{lamR}
|4\pi \cos k_0 /k_0^2|=\frac{1}{\rho l_B}+ \frac{1}{(\rho l_B)^2\rho}\int_0^{\Lambda} \frac{dk}{(2\pi)^2} \frac{2k^2}{4\pi \cos k /k^2+|4\pi \cos k_0 /k_0^2|}
\end{equation}
has no solutions for $1/(\rho l_B)>0$, because the integral diverges, and the instability line is shifted to $T=0$, as already discussed in Ref.~\cite{ciach:05:0} in the context of ions and in the original Brazovskii work~\cite{brazovskii:75:0} in the context of an order-parameter in the Landau-type theory for the order-disorder transition to an oscillatory state.

\begin{figure}[!htb]
    \includegraphics[scale=0.33]{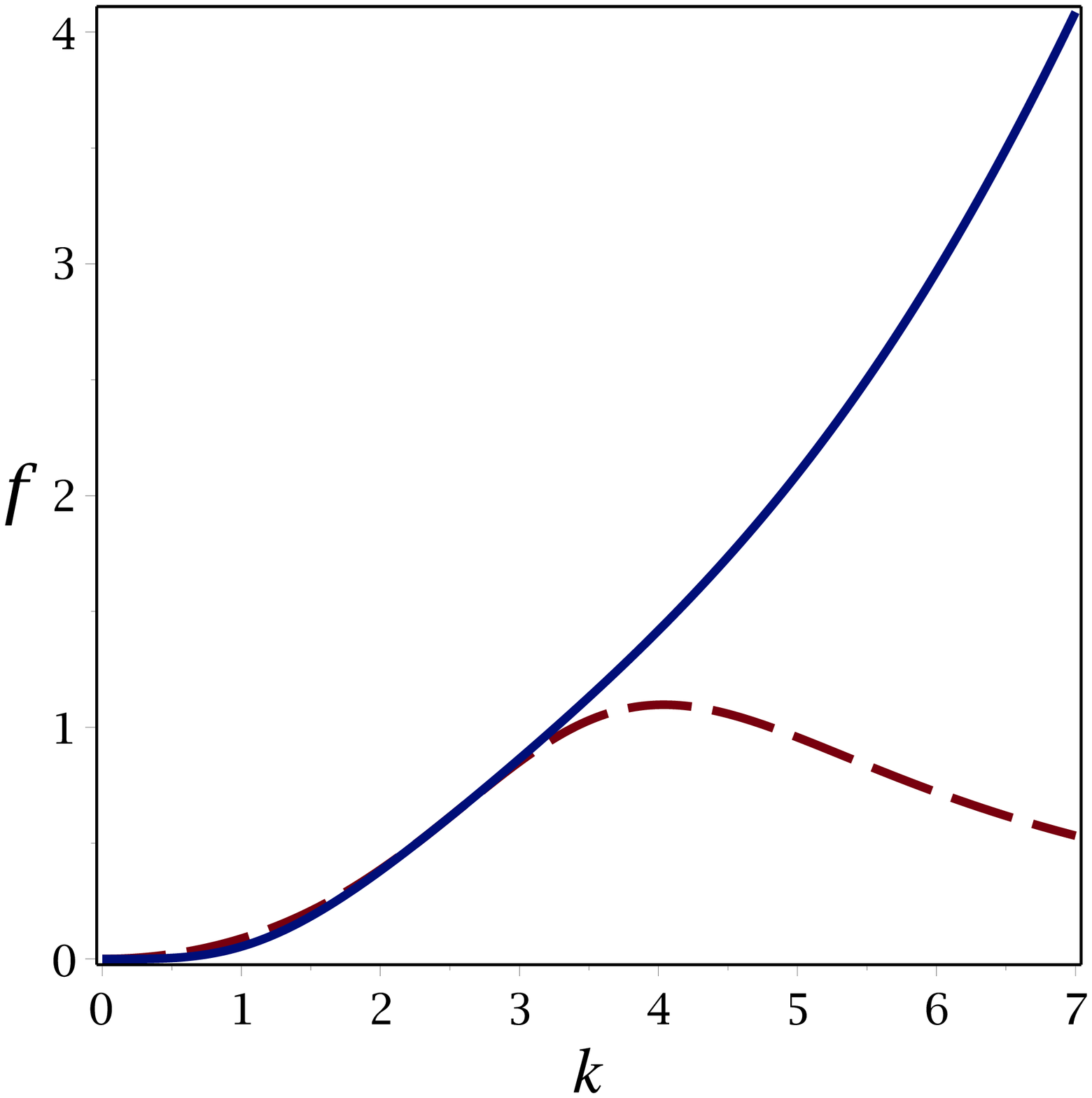}
    \includegraphics[scale=0.33]{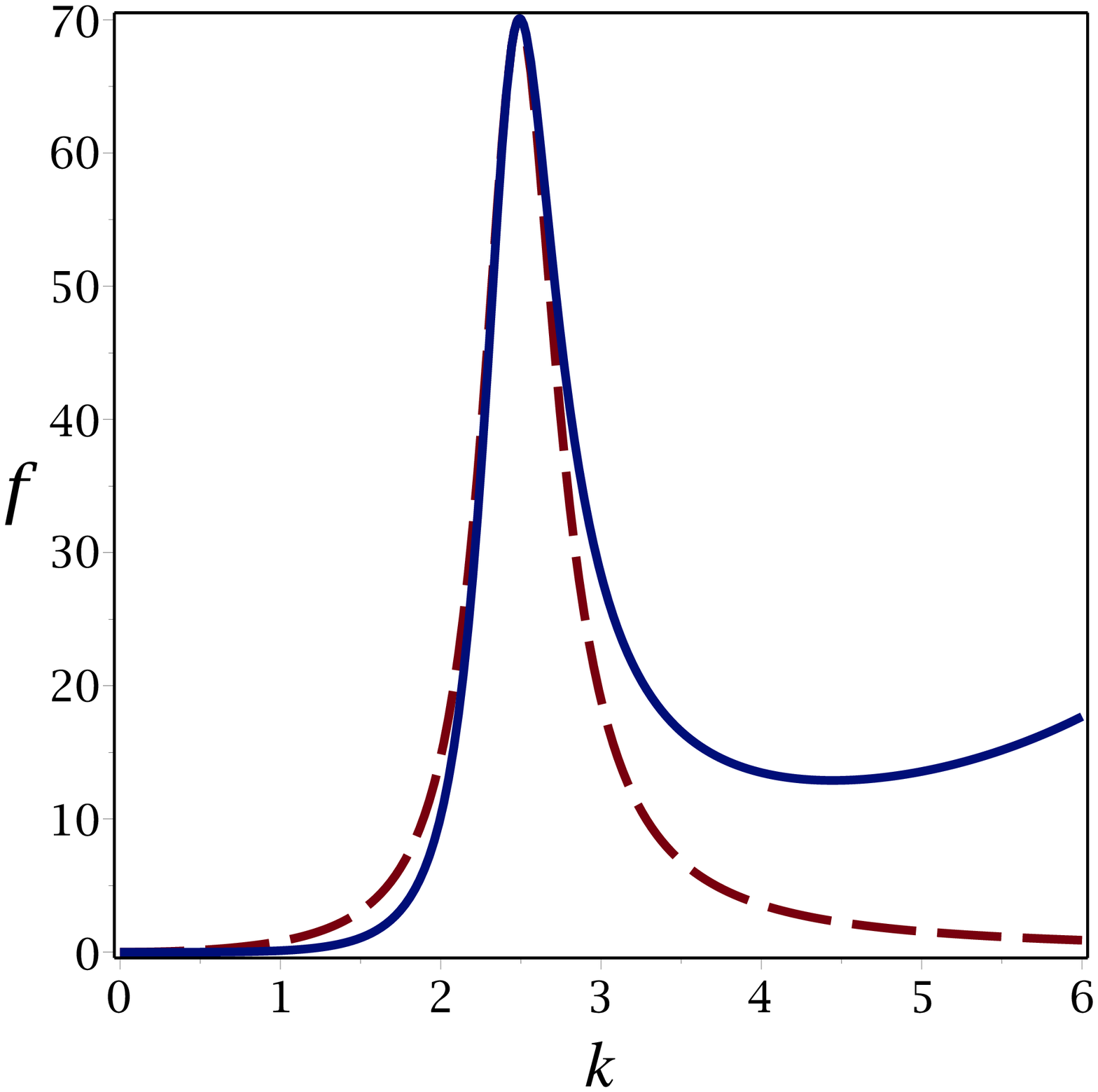}
\caption{The integrands, $f(k)=2k^2/\hat C_{cc}(k)$ (solid line), and   $f(k)=2k^2/\hat C_{a}(k)$ (dashed line),  on the LHS and the RHS  in Eq.~(\ref{Caa}), respectively.  Left panel: at the Kirkwood line ($S^R=11.8)$.  Right panel: for $S^R=1.7$  with $S^R$ defined in (\ref{KirR}).}
\label{fig:k_1_7}
\end{figure}

In Ref.~\cite{ciach:21:0},   $\hat C_{cc}(k)$  was approximated by 
\begin{equation}
\label{Ca}
 \hat C_{cc}(k)\approx   \hat C_{a}(k)=\hat C_{cc}(k_0)+\beta v (k^2-k_0^2)^2 ,
\end{equation}
obtained from a truncated Taylor expansion about the minimum at $k=k_0$, modified by the requirement that the approximation for  $\hat C_{cc}(k)$ should be an even function of $k$. Next it was argued that for relatively small values of $S^R$ corresponding to large $\rho l_B$, the cutoff-regularized integral can be approximated by
\begin{equation}
\label{Caa}
\int_0^{\pi} \frac{ dk}{(2\pi)^2} \frac{2k^2}{\hat C_{cc}(k)}\approx \int_0^{\infty} \frac{ dk}{(2\pi)^2} \frac{2k^2}{\hat C_{a}(k)},
\end{equation}
because in the case of a pronounced maximum of $\hat G_{cc}(k)$ at $k=k_0$, the main contribution to the integral comes from the vicinity of $k_0$, and
\begin{equation}
\int_{\pi}^{\infty} \frac{ dk}{(2\pi)^2} \frac{2k^2}{\hat C_{a}(k)}\ll \int_0^{\pi} \frac{ dk}{(2\pi)^2} \frac{2k^2}{\hat C_{a}(k)}.
\end{equation}
 The advantage of the approximation (\ref{Caa})  is the possibility of analytical calculation of the integral 
 on the RHS  of Eq.~(\ref{Caa}) and analytical solution of the corresponding approximate form of (\ref{Ccc3}) and (\ref{variance}).
As can be seen in Fig.~\ref{fig:k_1_7}, the integral  on the RHS  of Eq.~(\ref{Caa})
differs significantly from the integral 
on the LHS  of Eq.~(\ref{Caa}) at the Kirkwood line, and in general for large values of $S^R$, 
because the contribution to the integral from $k>\pi$ is large compared to the contribution from $k<\pi$. For small $S^R$, however, in particular for $S^R= 1.7$ shown in Fig.~\ref{fig:k_1_7}, the  contribution to the integral  $\int_0^{\infty} \frac{ dk}{(2\pi)^2} \frac{2k^2}{\hat C_{a}(k)}$ from $k>\pi$ is much smaller from the contribution from $k<\pi$. Thus, we can use the approximate expression (\ref{Ca}) for $\hat C_{cc}(k)$
only in the region of the  phase diagram limited to rather small $S^R$, i.e. to large densities and Bjerrum lengths. 
Under the assumption of large $l_B\rho$, the variance of the local charge density takes the form \cite{ciach:21:0} 
\begin{equation}
\label{<phi2>}
 \langle \phi^2\rangle\approx \frac{k_0}{4\pi\sqrt{\tilde C_{cc}(k_0)\beta v}}.
\end{equation}
$\tilde C_{cc}(k_0)$ is obtained by the self-consistent solution of (\ref{Ccc3}) and (\ref{<phi2>}). The parameters in (\ref{Gccr}) are  
 $ A_c\approx \langle \phi^2\rangle/k_0, \alpha_1\approx k_0$ and
\begin{equation}
\label{al0}
 \alpha_0^{-1}\approx8\pi \beta v\langle \phi^2\rangle\approx 1.1l_B \langle \phi^2\rangle.
\end{equation}
The proportionality of the decay length to $l_B\langle \phi^2\rangle$ agrees with our heuristic analysis of concentrated electrolytes. Moreover, both  the analytical results and our heuristic arguments are valid only when  $\rho l_B$ is large. 

\begin{figure}[!htb]
	\includegraphics[scale=0.4]{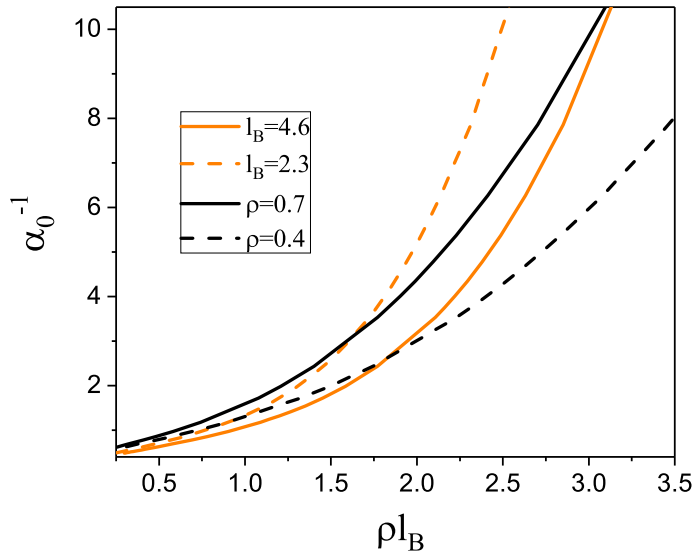}
	\caption{
		The correlation length of the oscillatory  decay of the charge-charge correlations, $\alpha_{0}^{-1}$, as a function of $\rho l_B$ for the fixed dimensionless density of ions $\rho$ and for the fixed Bjerrum length $l_B$ as it is given in the legend. $l_B$ is in units of the ion diameter $a$.
	}
	\label{fig:al0}
\end{figure}

\begin{figure}[htb]
	\includegraphics[scale=0.4]{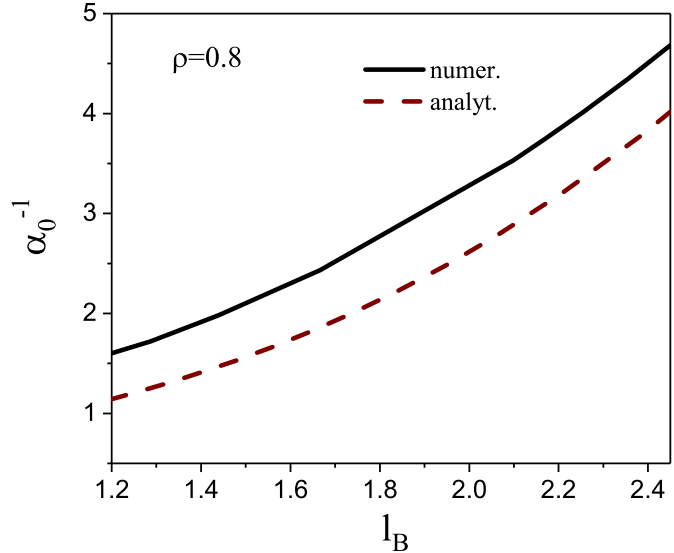}
	\caption{
		The correlation length of the oscillatory  decay of the charge-charge correlations $\alpha_{0}^{-1}$ as a function of the Bjerrum length $l_B$ for the dimensionless density of ions $\rho=0.8$ from the approximate analytical theory (bottom line) and with numerical solution of Eq.~(\ref{roR}) (top line). $l_B$ is in units of the ion diameter $a$.}
	\label{fig:al0_anal}
	\end{figure}

For a broader range of $\rho l_B$  we determine $\alpha_0$ numerically.
In Fig.~\ref{fig:al0}, the decay length $\alpha_0^{-1}$ is shown as a function of $\rho l_B$ for fixed $\rho$   and  fixed $l_B$, in particular, the results are given for $\rho=0.4, 0.7$ and  for $l_B=2.3,4.6$. 
Although in this theory $\alpha_0$ depends on a single thermodynamic parameter $S^R$ (see (\ref{poles})), $S^R$ depends on $\rho$ and $l_B$ separately (see (\ref{KirR})). For the fixed dimensionless densities of ions $\rho$, $\alpha_0^{-1}$ depends on $l_B\rho$ almost linearly for $\rho l_B>1.5$, but the slope increases with $\rho$. For the fixed Bjerrum lengths $l_B$, $\alpha_0^{-1}$ also increases almost linearly  with  $l_B\rho$ but it occurs for larger   $\rho l_B$ and  the slope decreases when $l_B$ increases. 
It turns out that our analytical solution underestimates a little  the correlation length $\alpha_0^{-1}$, but the lines $\alpha_0^{-1}(l_B)$ obtained  with and without the approximation (\ref{Ca}) are almost parallel, and shifted with respect to each other by $\sim 0.5$  (see Fig.~\ref{fig:al0_anal}). 

\begin{figure}[!htb]
	\begin{center}
		\includegraphics[scale=0.35]{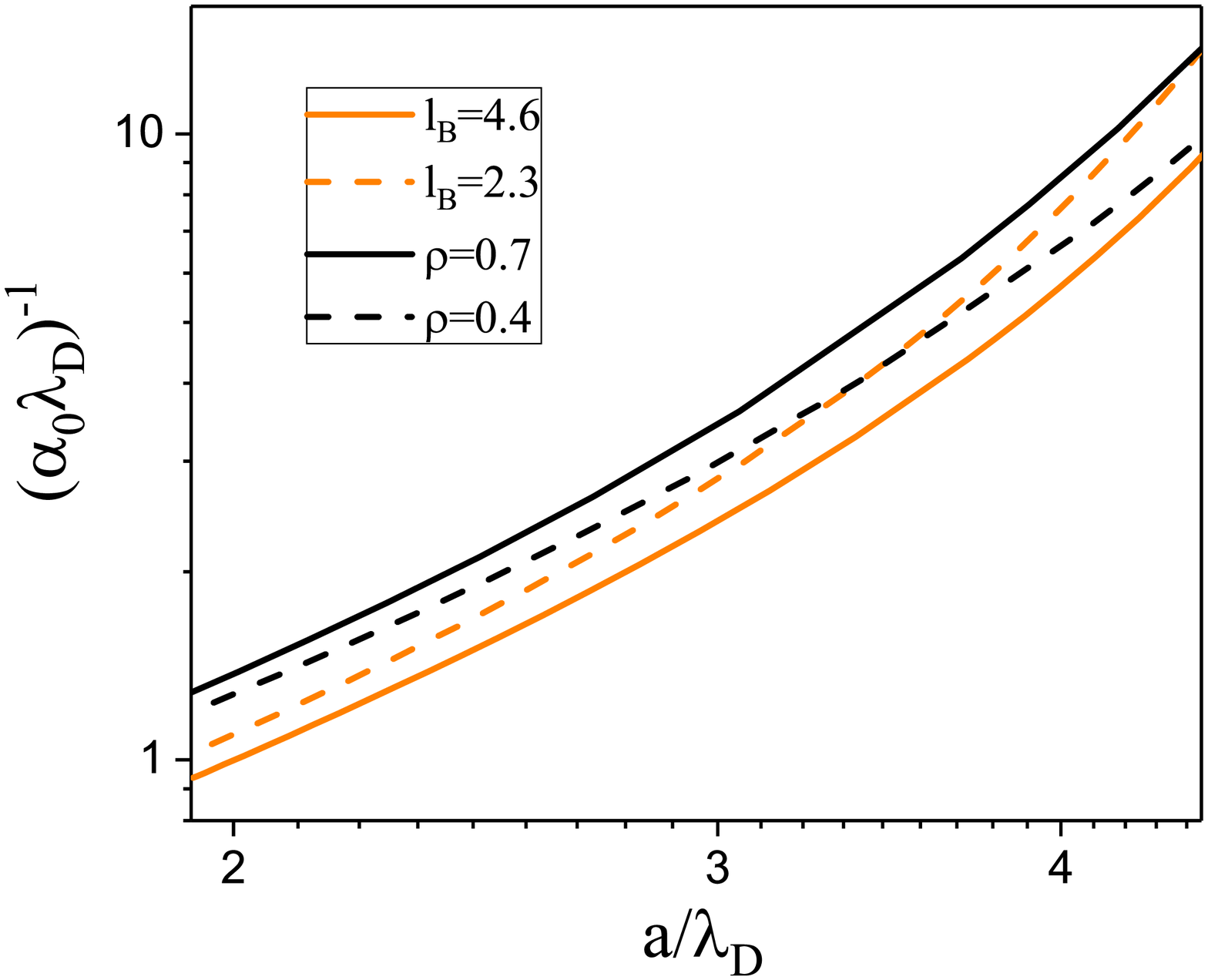}
	\includegraphics[scale=0.35]{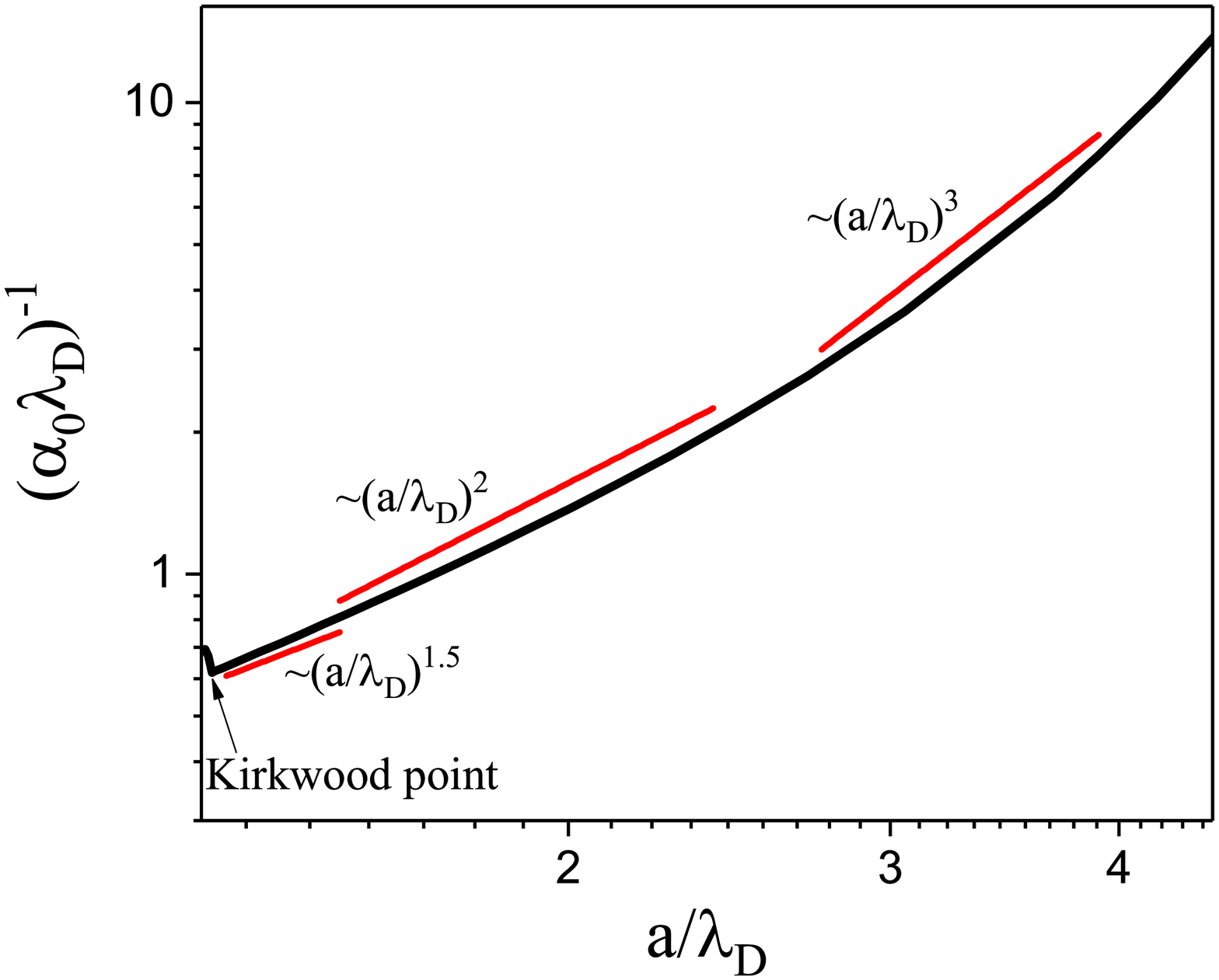}
		\caption{
			Left panel: 
			A double logarithmic plot of  the ratio of the correlation length of the oscillatory decay of the charge-charge correlations and Debye length, $(\alpha_{0}\lambda_{D})^{-1}$, as a function of the inverse of the Debye length, $a/\lambda_{D}$,  for the dimensionless density of ions $\rho$  and the Bjerrum length $l_B$ as it is given in the legend. Right panel: the same as in the left panel for $\rho=0.7$ with the indication of the regions where different scaling regimes hold. The arrow points to the cusp (Kirkwood point) where  crossover from monotonic to oscillatory decay occurs.
		}\label{fig:al0_log1}
	\end{center}
\end{figure}

Figure~\ref{fig:al0_log1} (left panel) shows  the ratio of the charge-charge decay length to the Debye length,  $\alpha_{0}^{-1}/\lambda_{D}$,  as a function  of the inverse of the Debye length, $a/\lambda_{D}$ (in log-log scale).  The results are given for  the same values of fixed $\rho$ and $l_B$ as in Fig.~\ref{fig:al0}.   
Using these results,  we test the scaling relation $\alpha_{0}^{-1}/\lambda_{D}\sim (a/\lambda_{D})^n$. Our analysis revealed the existence of
different scaling regimes corresponding to different values of $n$.  In particular, we found that $n=3$ for the range $2.5<a/\lambda_D<4$, $n=2$ for  $1.5<a/\lambda_{D}<2.5$, and $n=1.5$ for $a/\lambda_{D}$ closer to  the Kirkwood point.  A typical picture of the scaling regimes is presented in Fig.~\ref{fig:al0_log1} (right panel) for the case  $\rho=0.7$  (see Supplemental Material for details).

In Fig.~\ref{fig:al0_exp}, we present the comparison of  the predictions of the mesoscopic theory  with the experimental data for aqueous NaCl solutions \cite{smith:16:0}.
As it is seen, we obtain a semiquantitative agreement with experimental findings for  $2<a/\lambda_{D}<4$. Moreover, the agreement becomes quantitative if one assumes that the average diameter of hydrated ions in the experiment is equal to $0.67$ nm.

\begin{figure}[!htb]
	\begin{center}
		\includegraphics[scale=0.42]{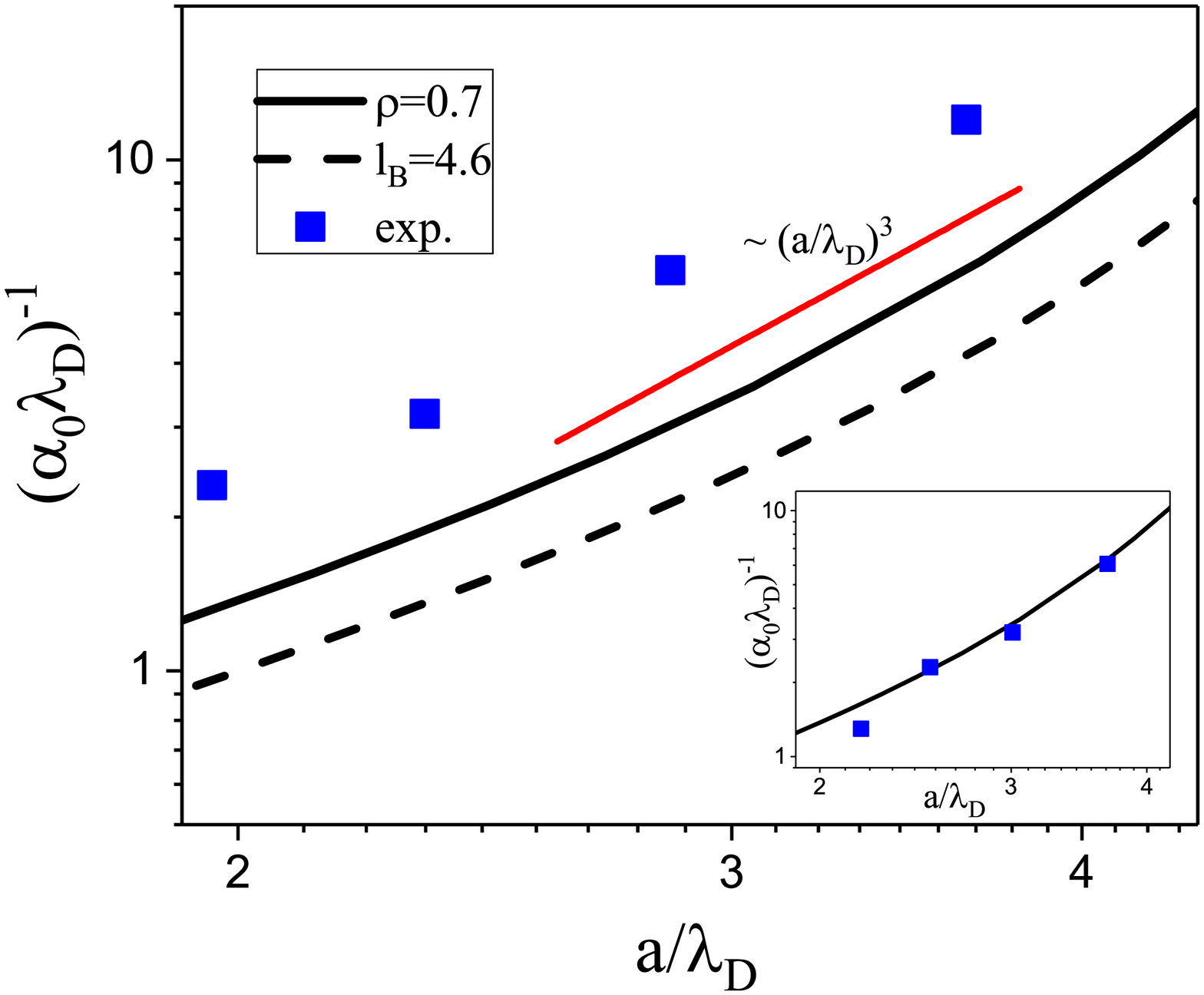}
		\caption{ A double logarithmic plot of the ratio $\alpha_0^{-1}/\lambda_D$ as a function of $a/\lambda_D$. The blue solid squares are the experimental results of Smith et al. \cite{smith:16:0} (Supporting Information) for NaCl in water for $a=0.52$ nm (hydrated ions). The curves show the theoretical predictions for the dimensionless density of ions $\rho=0.7$ (solid line) and for the Bjerrum length $l_B=4.6$ (dashed line). The red solid line depicts power law as noted. $l_B$ is in units of the ion diameter $a$. Inset: the same as in the main plot  for $\rho=0.7$, assuming that in the experiment the  hydrated ion diameter was $0.67$ nm.
		}\label{fig:al0_exp}
	\end{center}
\end{figure}

\subsection{Asymptotic decay of density-density correlations}
\label{sec:corfuro}
\begin{figure}[!htb]
	\begin{center}
		\includegraphics[scale=0.35]{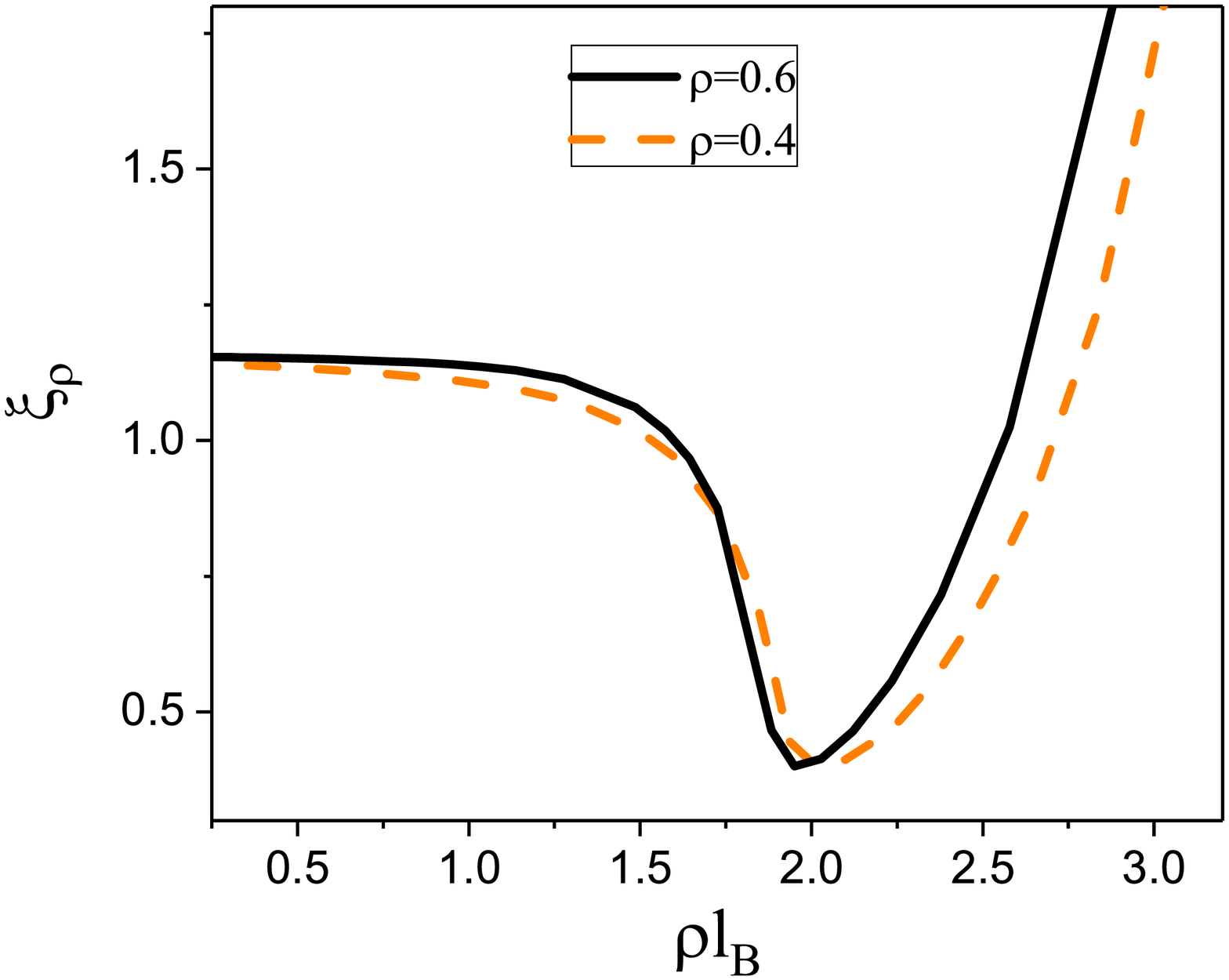}
		\includegraphics[scale=0.35]{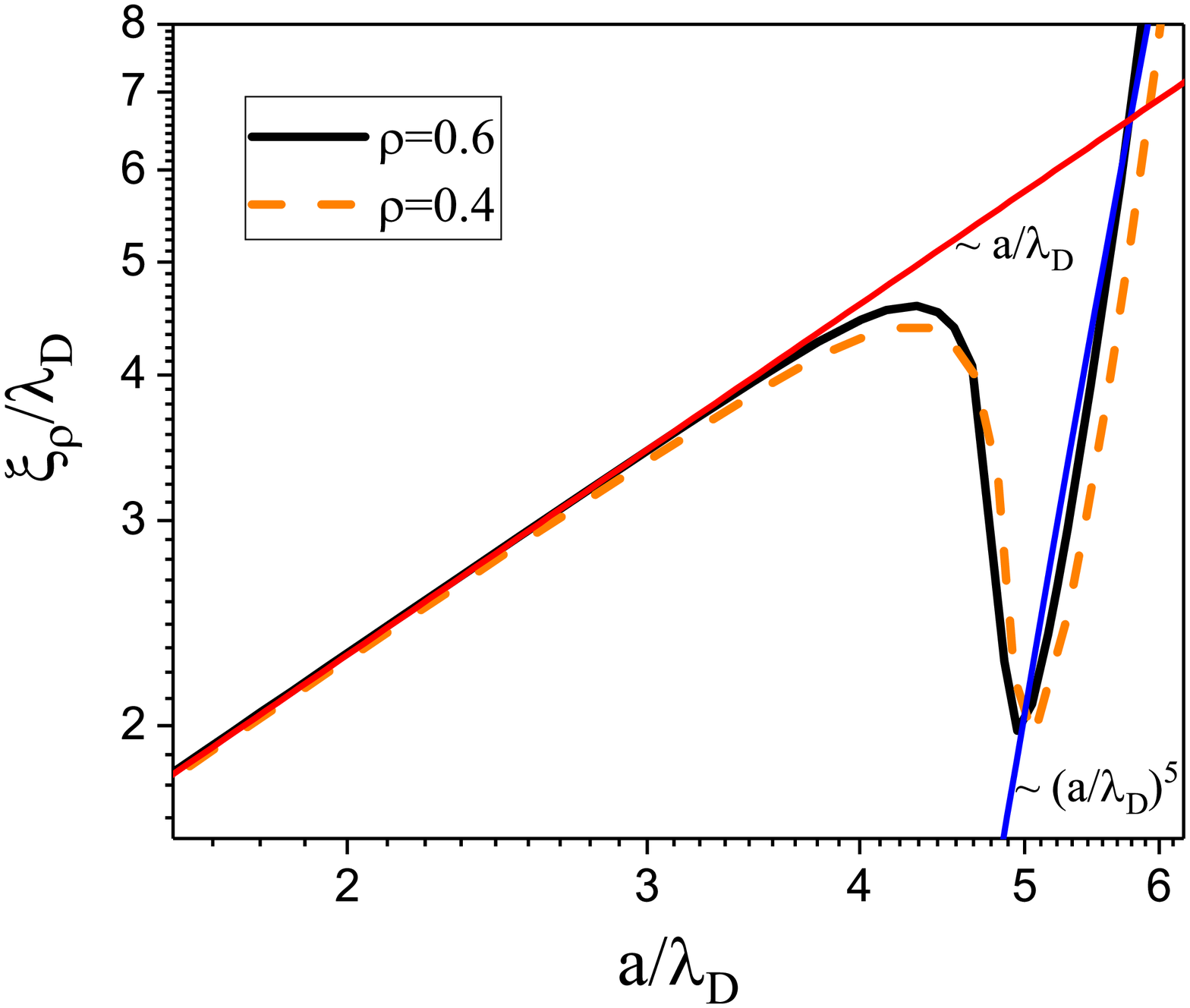}
		\caption{
			Left panel: the decay length of the density-density correlations  $\xi_{\rho}$ as a function of $\rho l_B$ for the fixed dimensionless densities of ions $\rho=0.4$ (dashed line) and $\rho=0.6$ (solid line). $l_B$ is in units of the ion diameter $a$. Right panel: 
			a double logarithmic plot of  the ratio of the decay length of the density-density correlation function and Debye length, $\xi_{\rho}/\lambda_{D}$, as a function of the inverse of the Debye length, $a/\lambda_{D}$,  for  the dimensionless densities of ions $\rho=0.4$ (dashed line) and $\rho=0.6$ (solid line). Red and blue solid lines depict  power laws as noted.
		}\label{fig:xi_1}
	\end{center}
\end{figure}

\begin{figure}[!htb]
	\begin{center}
		\includegraphics[scale=0.35]{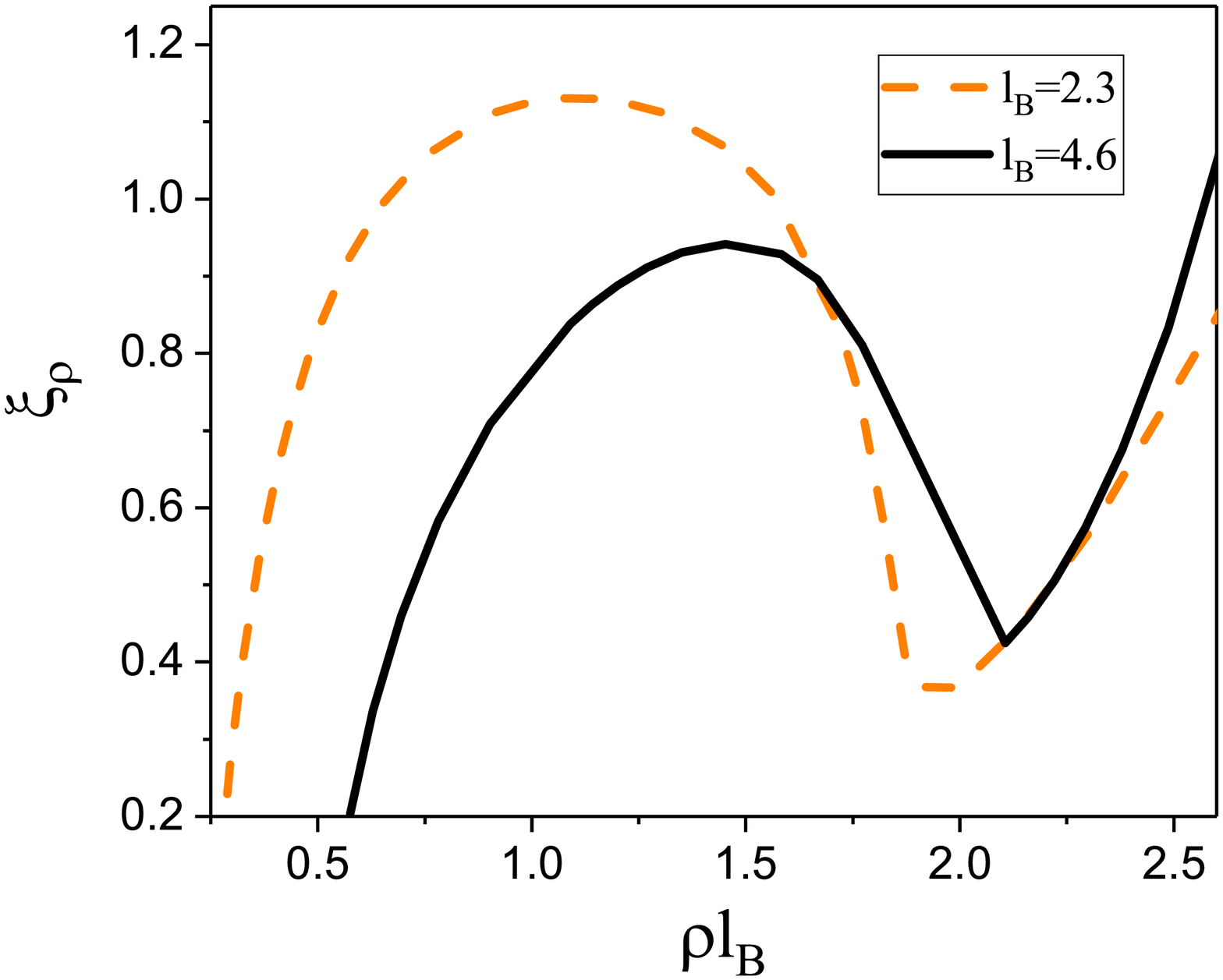}
		\includegraphics[scale=0.35]{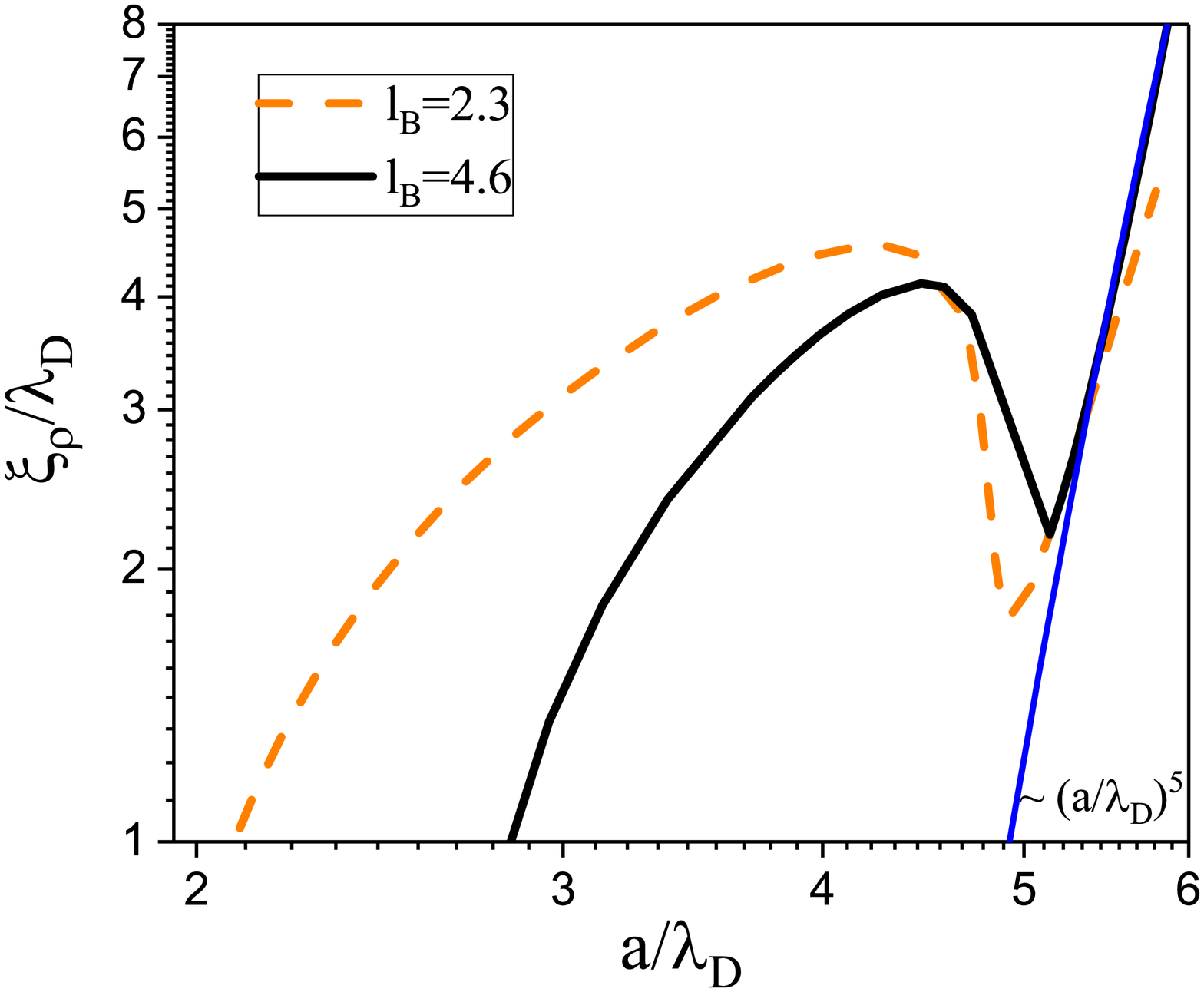}
		\caption{Left panel: the decay length of the density-density correlations  $\xi_{\rho}$ as a function of $\rho l_B$ for the fixed  Bjerrum lengths $l_B=2.3$ and $l_B=4.6$. $l_B$ is in units of the ion diameter $a$.
			Right panel: a double logarithmic plot of  the ratio of the decay length of the density-density correlations and Debye length, $\xi_{\rho}/\lambda_{D}$, as a function of the inverse of the Debye length, $a/\lambda_{D}$,  for the Bjerrum lengths $l_B=2.3$ (dashed line) and $l_B=4.6$ (solid line). Blue solid line depicts a power law as noted.
		}\label{fig:xi_2}
	\end{center}
\end{figure}

  When the fluctuation contribution is not included in (\ref{Crr3}), the density-density correlations are strictly short-range in our local-density approximation when the energy depends only on the charge density. The $k$-dependence of $\hat C_{\rho\rho}(k)$ comes from  $\hat V_{fl}(k)$ that is induced by the correlations between fluctuations.  $\hat V_{fl}(k)$ takes a negative minimum for $k=0$ (see (\ref{Vfl})), and for $k\to 0$ can be Taylor expanded. The Taylor expansion of  $\hat V_{fl}(k)$ takes the same mathematical form as the Taylor expanded attractive interactions in Fourier representation, and we have the approximation
\begin{equation}
\label{Te}
\hat C_{\rho\rho}(k)=R_0+R_2k^2 +O(k^4).
\end{equation} 
The rather lengthy formulas for $R_n$ can be found in Ref.~\cite{patsahan:22:0} (Appendix~A).
From (\ref{Te}) we obtain the asymptotic decay in real space for  $r\gg 1$,
\begin{equation}
\label{Grr}
 G_{\rho\rho}(r)=\frac{1}{4\pi R_2 }\frac{\exp(-r\sqrt{R_0/R_2})}{r}.
\end{equation}
We can see that the asymptotic decay of the density-density correlations is monotonic, with the decay length $\xi_{\rho}=\sqrt{R_2/R_0}$ depending on the charge-charge correlations (see (\ref{Vfl}) and (\ref{Crr3})).
Explicit expressions for $R_0,R_2$ can be obtained by a self-consistent solution of  (\ref{Te}), (\ref{Crr3})  and (\ref{Vfl}) with Taylor-expanded $\hat V_{fl}(k)$. $\hat V_{fl}(k)$ depends on  the form of $G_{cc}(r)$ that except from large densities must be determined numerically, therefore the self-consistent solution of (\ref{Te}), (\ref{Crr3}) and (\ref{Vfl}) is obtained numerically too. We assume the Carnahan-Starling approximation~\cite{Mansoori:71},
\begin{equation} 
\beta f_{ex}(\zeta)=
\rho\Big[
\frac{4\zeta-3\zeta^2}{(1-\zeta)^2}-1
\Big],
\end{equation}
 with $\zeta=\pi\rho/6$,
and show the resulting correlation length in Fig.~\ref{fig:xi_1}  and in Fig.~\ref{fig:xi_2} for fixed  $\rho$ and $l_B$, respectively. In both figures, the left panel shows the dependence of the decay length $\xi_{\rho}$ on $\rho l_B$. In the right  panel,  we show the ratio  $\xi_{\rho}/\lambda_{D}$ as a function of the inverse of the Debye length $a/\lambda_{D}$ in log-log scale. For both, the fixed $\rho$  and the fixed $l_B$,  $\xi_{\rho}$ ($\xi_{\rho}/\lambda_{D}$) is an increasing  function of $\rho l_B$ ($a/\lambda_{D}$)  for $\rho l_B> 2$ ($a/\lambda_D> 5$) and its behaviour is nonmonotonic  with  a rather deep minimum for $\rho l_B\approx 2$ ($a/\lambda_D\approx 5$).  Furthermore, $\xi_{\rho}/\lambda_{D}$ increases as $(a/\lambda_{D})^5$ for $a/\lambda_D> 5$.  For  $\rho l_B< 2$  ($a/\lambda_D< 5$), the behaviour of $\xi_{\rho}$ for the fixed $\rho$ and the fixed $l_B$ is completely different. In this case, $\xi_{\rho}$ ($\xi_{\rho}/\lambda_{D}$) decreases (increases) when $\rho l_B$ ($a/\lambda_{D}$) increases for the fixed dimensionless densities of ions $\rho=0.4$ and $\rho=0.6$. Interestingly,  in this region $\xi_{\rho}/\lambda_{D}$  increases linearly with  $a/\lambda_{D}$. For the fixed Bjerrum lengths  $l_B=2.3$ and $l_B=4.6$, $\xi_{\rho}$ and $\xi_{\rho}/\lambda_{D}$ shows a nonmonotonic behaviour with a maximum and the maximum  is higher for $l_B=2.3$.  Simultaneously, for $l_B=4.6$, both  the maximum and the minimum  are shifted to larger  $\rho l_B$  ($a/\lambda_D$) when compared with  $l_B=2.3$.

\section{Conclusions}
\label{sec:conclusions}

We have found that the dependence of the correlation length in electrolytes on the density of ions and the Bjerrum length has a more complex behavior than just a simple scaling $\lambda_s/\lambda_D=\alpha_0^{-1}/\lambda_D\sim (a/\lambda_D)^n$. The above formula can be a fair approximation for the charge-charge correlation length, but with the exponent $n$ taking different values for different ranges of $a/\lambda_D$. From our theory it follows that $n$ increases with increasing $a/\lambda_D$.  In particular, for  the fixed dimensionless density of ions $\rho=0.7$, we find  $n=3$ for the range $2.5<a/\lambda_D<4$, $n=2$ for the range $1.5<a/\lambda_{D}<2.5$, and $n=1.5$ for $a/\lambda_D$ closer to the Kirkwood point. 
The existence of different scaling  regimes for different ranges of $a/\lambda_D$  was already reported  in \cite{Outhwaite2021}. 
It should be noted that  the scaling exponent $n\approx 2$  was found  for the RPM theoretically \cite{Adar2019}, as well as in simulations \cite{Zeman2021,KruckerVelasquez2021}.
Moreover,  the anomalous underscreening for the RPM that agrees with experimental results  was obtained very recently in simulations  and by theory accounting for ions pairing in Ref.~\cite{arxiv.2209.03486}. 

 We should note that  $n=3$ perfectly fits the experimental results for $a/\lambda_D>2$, but for $a/\lambda_D<2$, the scaling  $\lambda_s/\lambda_D\sim (a/\lambda_D)^3$ is less good.
 Notably, $n=3$ corresponds to $\lambda_s\sim l_B\rho$. In our theory, we obtain $\lambda_s\sim l_B\langle\phi^2\rangle$ for large $l_B\rho$, and the scaling $\lambda_s\sim l_B\rho$ is obtained when the variance of the local charge behaves as $ \langle\phi^2\rangle\propto \rho$, which is a rough approximation. The relation $\lambda_s\sim l_B\langle\phi^2\rangle$ is also obtained from the very simplified lattice model, with the charge taking the value $\pm\sqrt\langle\phi^2\rangle$ in the lattice cells with the lattice constant $a$. 
Both, the simplified model and our analytical theory are not valid for medium density of ions, and from numerical solution of our equations  we obtain the exponent $n$ in fair agreement with simulations and other theoretical results.
 We should also mention quite good agreement between the Kirkwood line obtained in our theory and in simulations~\cite{Cats2021}. 
 We conclude that the results of our theory form a bridge between the experiment, simulations and other theoretical predictions.
 
 In addition to the charge-charge correlation we study the density-density correlations. The obtained   density-density correlation function  decays monotonically. The corresponding asymptotic decay length $\xi_{\rho}$ depends on the charge-charge correlations in addition to  the correlations associated with the hard spheres (steric) interactions. This result significantly differs from  the results  obtained previously for the decay length of the density-density correlation function which {take into account only the contribution from the steric interactions \cite{Coupette2018,Cats2021}}. 

\bibliography{bibliography_22.bib}

\begin{thebibliography}{27}
\expandafter\ifx\csname natexlab\endcsname\relax\def\natexlab#1{#1}\fi
\expandafter\ifx\csname bibnamefont\endcsname\relax
  \def\bibnamefont#1{#1}\fi
\expandafter\ifx\csname bibfnamefont\endcsname\relax
  \def\bibfnamefont#1{#1}\fi
\expandafter\ifx\csname citenamefont\endcsname\relax
  \def\citenamefont#1{#1}\fi
\expandafter\ifx\csname url\endcsname\relax
  \def\url#1{\texttt{#1}}\fi
\expandafter\ifx\csname urlprefix\endcsname\relax\def\urlprefix{URL }\fi
\providecommand{\bibinfo}[2]{#2}
\providecommand{\eprint}[2][]{\url{#2}}

\bibitem[{\citenamefont{Torquato}(2018)}]{torquato:18:0}
\bibinfo{author}{\bibfnamefont{S.}~\bibnamefont{Torquato}},
  \bibinfo{journal}{{\it Physics Reports}} \textbf{\bibinfo{volume}{745}},
  \bibinfo{pages}{1} (\bibinfo{year}{2018}).

\bibitem[{\citenamefont{Smith et~al.}(2016)\citenamefont{Smith, Lee, and
  Perkin}}]{smith:16:0}
\bibinfo{author}{\bibfnamefont{A.~M.} \bibnamefont{Smith}},
  \bibinfo{author}{\bibfnamefont{A.~A.} \bibnamefont{Lee}}, \bibnamefont{and}
  \bibinfo{author}{\bibfnamefont{S.}~\bibnamefont{Perkin}},
  \bibinfo{journal}{{\it J. Phys. Chem. Lett.}} \textbf{\bibinfo{volume}{7}},
  \bibinfo{pages}{2157} (\bibinfo{year}{2016}).

\bibitem[{\citenamefont{Lee et~al.}(2017)\citenamefont{Lee, Perez-Martinez,
  Smith, and Perkin}}]{lee:17:0}
\bibinfo{author}{\bibfnamefont{A.}~\bibnamefont{Lee}},
  \bibinfo{author}{\bibfnamefont{C.~S.} \bibnamefont{Perez-Martinez}},
  \bibinfo{author}{\bibfnamefont{A.~M.} \bibnamefont{Smith}}, \bibnamefont{and}
  \bibinfo{author}{\bibfnamefont{S.}~\bibnamefont{Perkin}},
  \bibinfo{journal}{{\it Phys. Rev. Lett.}} \textbf{\bibinfo{volume}{119}},
  \bibinfo{pages}{026002} (\bibinfo{year}{2017}).

\bibitem[{\citenamefont{Groves et~al.}(2021)\citenamefont{Groves,
  Perez-Martinez, Lhermerout, and Perkin}}]{Groves2021}
\bibinfo{author}{\bibfnamefont{T.~S.} \bibnamefont{Groves}},
  \bibinfo{author}{\bibfnamefont{C.~S.} \bibnamefont{Perez-Martinez}},
  \bibinfo{author}{\bibfnamefont{R.}~\bibnamefont{Lhermerout}},
  \bibnamefont{and} \bibinfo{author}{\bibfnamefont{S.}~\bibnamefont{Perkin}},
  \bibinfo{journal}{{\it J. Phys. Chem. Lett.}} \textbf{\bibinfo{volume}{12}},
  \bibinfo{pages}{1702} (\bibinfo{year}{2021}).

\bibitem[{\citenamefont{Zeman et~al.}(2020)\citenamefont{Zeman, Kondrat, and
  Holm}}]{Zeman2020}
\bibinfo{author}{\bibfnamefont{J.}~\bibnamefont{Zeman}},
  \bibinfo{author}{\bibfnamefont{S.}~\bibnamefont{Kondrat}}, \bibnamefont{and}
  \bibinfo{author}{\bibfnamefont{C.}~\bibnamefont{Holm}},
  \bibinfo{journal}{{\it Chem. Comm.}} \textbf{\bibinfo{volume}{56}},
  \bibinfo{pages}{15635} (\bibinfo{year}{2020}).

\bibitem[{\citenamefont{Adar et~al.}(2019)\citenamefont{Adar, Safran, Diamant,
  and Andelman}}]{Adar2019}
\bibinfo{author}{\bibfnamefont{R.~M.} \bibnamefont{Adar}},
  \bibinfo{author}{\bibfnamefont{S.~A.} \bibnamefont{Safran}},
  \bibinfo{author}{\bibfnamefont{H.}~\bibnamefont{Diamant}}, \bibnamefont{and}
  \bibinfo{author}{\bibfnamefont{D.}~\bibnamefont{Andelman}},
  \bibinfo{journal}{{\it Phys. Rev. E}} \textbf{\bibinfo{volume}{100}},
  \bibinfo{pages}{042615} (\bibinfo{year}{2019}).

\bibitem[{\citenamefont{Ciach and Patsahan}(2021)}]{ciach:21:0}
\bibinfo{author}{\bibfnamefont{A.}~\bibnamefont{Ciach}} \bibnamefont{and}
  \bibinfo{author}{\bibfnamefont{O.}~\bibnamefont{Patsahan}},
  \bibinfo{journal}{{\it J. Phys.: Condens. Matter}}
  \textbf{\bibinfo{volume}{33}}, \bibinfo{pages}{37LT01}
  (\bibinfo{year}{2021}).

\bibitem[{\citenamefont{Ciach}(2008)}]{ciach:08:1}
\bibinfo{author}{\bibfnamefont{A.}~\bibnamefont{Ciach}}, \bibinfo{journal}{{\it
  Phys. Rev. E}} \textbf{\bibinfo{volume}{78}}, \bibinfo{pages}{061505}
  (\bibinfo{year}{2008}).

\bibitem[{\citenamefont{Patsahan et~al.}(2022)\citenamefont{Patsahan, Meyra,
  and Ciach}}]{patsahan:22:0}
\bibinfo{author}{\bibfnamefont{O.}~\bibnamefont{Patsahan}},
  \bibinfo{author}{\bibfnamefont{A.}~\bibnamefont{Meyra}}, \bibnamefont{and}
  \bibinfo{author}{\bibfnamefont{A.}~\bibnamefont{Ciach}},
  \bibinfo{journal}{{\it J. Mol. Liq.}} \textbf{\bibinfo{volume}{363}},
  \bibinfo{pages}{119844} (\bibinfo{year}{2022}).

\bibitem[{\citenamefont{Goodwin and Kornyshev}(2017)}]{Goodwin2017}
\bibinfo{author}{\bibfnamefont{Z.~A.} \bibnamefont{Goodwin}} \bibnamefont{and}
  \bibinfo{author}{\bibfnamefont{A.~A.} \bibnamefont{Kornyshev}},
  \bibinfo{journal}{{\it Electrochem. Commun.}} \textbf{\bibinfo{volume}{82}},
  \bibinfo{pages}{129} (\bibinfo{year}{2017}).

\bibitem[{\citenamefont{Ludwig et~al.}(2018)\citenamefont{Ludwig, Dasbiswas,
  Talapin, and Vaikuntanathan}}]{Ludwig2018}
\bibinfo{author}{\bibfnamefont{N.~B.} \bibnamefont{Ludwig}},
  \bibinfo{author}{\bibfnamefont{K.}~\bibnamefont{Dasbiswas}},
  \bibinfo{author}{\bibfnamefont{D.~V.} \bibnamefont{Talapin}},
  \bibnamefont{and}
  \bibinfo{author}{\bibfnamefont{S.}~\bibnamefont{Vaikuntanathan}},
  \bibinfo{journal}{{\it J. Chem. Phys.}} \textbf{\bibinfo{volume}{149}},
  \bibinfo{pages}{164505} (\bibinfo{year}{2018}).

\bibitem[{\citenamefont{Coupette et~al.}(2018)\citenamefont{Coupette, Lee, and
  H\"{a}rtel}}]{Coupette2018}
\bibinfo{author}{\bibfnamefont{F.}~\bibnamefont{Coupette}},
  \bibinfo{author}{\bibfnamefont{A.~A.} \bibnamefont{Lee}}, \bibnamefont{and}
  \bibinfo{author}{\bibfnamefont{A.}~\bibnamefont{H\"{a}rtel}},
  \bibinfo{journal}{{\it Phys. Rev. Lett.}} \textbf{\bibinfo{volume}{121}}
  (\bibinfo{year}{2018}).

\bibitem[{\citenamefont{Rotenberg et~al.}(2018)\citenamefont{Rotenberg,
  Bernard, and Hansen}}]{Rotenberg_2018}
\bibinfo{author}{\bibfnamefont{B.}~\bibnamefont{Rotenberg}},
  \bibinfo{author}{\bibfnamefont{O.}~\bibnamefont{Bernard}}, \bibnamefont{and}
  \bibinfo{author}{\bibfnamefont{J.-P.} \bibnamefont{Hansen}},
  \bibinfo{journal}{{\it J. Phys.: Condens. Matter}}
  \textbf{\bibinfo{volume}{30}}, \bibinfo{pages}{054005}
  (\bibinfo{year}{2018}).

\bibitem[{\citenamefont{Coles et~al.}(2020)\citenamefont{Coles, Park, Nikam,
  Kandu\v{c}, Dzubiella, and Rotenberg}}]{Coles2020}
\bibinfo{author}{\bibfnamefont{S.~W.} \bibnamefont{Coles}},
  \bibinfo{author}{\bibfnamefont{C.}~\bibnamefont{Park}},
  \bibinfo{author}{\bibfnamefont{R.}~\bibnamefont{Nikam}},
  \bibinfo{author}{\bibfnamefont{M.}~\bibnamefont{Kandu\v{c}}},
  \bibinfo{author}{\bibfnamefont{J.}~\bibnamefont{Dzubiella}},
  \bibnamefont{and}
  \bibinfo{author}{\bibfnamefont{B.}~\bibnamefont{Rotenberg}},
  \bibinfo{journal}{{\it J. Phys. Chem. B}} \textbf{\bibinfo{volume}{124}},
  \bibinfo{pages}{1778} (\bibinfo{year}{2020}).

\bibitem[{\citenamefont{Cats et~al.}(2021)\citenamefont{Cats, Evans,
  H\"{a}rtel, and van Roij}}]{Cats2021}
\bibinfo{author}{\bibfnamefont{P.}~\bibnamefont{Cats}},
  \bibinfo{author}{\bibfnamefont{R.}~\bibnamefont{Evans}},
  \bibinfo{author}{\bibfnamefont{A.}~\bibnamefont{H\"{a}rtel}},
  \bibnamefont{and} \bibinfo{author}{\bibfnamefont{R.}~\bibnamefont{van Roij}},
  \bibinfo{journal}{{\it J. Chem. Phys.}} \textbf{\bibinfo{volume}{154}},
  \bibinfo{pages}{124504} (\bibinfo{year}{2021}).

\bibitem[{\citenamefont{Zeman et~al.}(2021)\citenamefont{Zeman, Kondrat, and
  Holm}}]{Zeman2021}
\bibinfo{author}{\bibfnamefont{J.}~\bibnamefont{Zeman}},
  \bibinfo{author}{\bibfnamefont{S.}~\bibnamefont{Kondrat}}, \bibnamefont{and}
  \bibinfo{author}{\bibfnamefont{C.}~\bibnamefont{Holm}},
  \bibinfo{journal}{{\it J. Chem. Phys.}} \textbf{\bibinfo{volume}{155}},
  \bibinfo{pages}{204501} (\bibinfo{year}{2021}).

\bibitem[{\citenamefont{Outhwaite and Bhuiyan}(2021)}]{Outhwaite2021}
\bibinfo{author}{\bibfnamefont{C.~W.} \bibnamefont{Outhwaite}}
  \bibnamefont{and} \bibinfo{author}{\bibfnamefont{L.~B.}
  \bibnamefont{Bhuiyan}}, \bibinfo{journal}{{\it J. Chem. Phys.}}
  \textbf{\bibinfo{volume}{155}}, \bibinfo{pages}{014504}
  (\bibinfo{year}{2021}).

\bibitem[{\citenamefont{Krucker-Velasquez and
  Swan}(2021)}]{KruckerVelasquez2021}
\bibinfo{author}{\bibfnamefont{E.}~\bibnamefont{Krucker-Velasquez}}
  \bibnamefont{and} \bibinfo{author}{\bibfnamefont{J.~W.} \bibnamefont{Swan}},
  \bibinfo{journal}{{\it J. Chem. Phys.}} \textbf{\bibinfo{volume}{155}},
  \bibinfo{pages}{134903} (\bibinfo{year}{2021}).

\bibitem[{\citenamefont{Kumar et~al.}(2022)\citenamefont{Kumar, Cats, Alotaibi,
  Ayirala, Yousef, van Roij, Siretanu, and Mugele}}]{Kumar2022}
\bibinfo{author}{\bibfnamefont{S.}~\bibnamefont{Kumar}},
  \bibinfo{author}{\bibfnamefont{P.}~\bibnamefont{Cats}},
  \bibinfo{author}{\bibfnamefont{M.~B.} \bibnamefont{Alotaibi}},
  \bibinfo{author}{\bibfnamefont{S.~C.} \bibnamefont{Ayirala}},
  \bibinfo{author}{\bibfnamefont{A.~A.} \bibnamefont{Yousef}},
  \bibinfo{author}{\bibfnamefont{R.}~\bibnamefont{van Roij}},
  \bibinfo{author}{\bibfnamefont{I.}~\bibnamefont{Siretanu}}, \bibnamefont{and}
  \bibinfo{author}{\bibfnamefont{F.}~\bibnamefont{Mugele}},
  \bibinfo{journal}{{\it Journal of Colloid and Interface Science}}
  \textbf{\bibinfo{volume}{622}}, \bibinfo{pages}{819} (\bibinfo{year}{2022}).

\bibitem[{\citenamefont{H\"{a}rtel et~al.}(2022)\citenamefont{H\"{a}rtel,
  B\"{u}ltmann, and Coupette}}]{arxiv.2209.03486}
\bibinfo{author}{\bibfnamefont{A.}~\bibnamefont{H\"{a}rtel}},
  \bibinfo{author}{\bibfnamefont{M.}~\bibnamefont{B\"{u}ltmann}},
  \bibnamefont{and} \bibinfo{author}{\bibfnamefont{F.}~\bibnamefont{Coupette}},
  \emph{\bibinfo{title}{Anomalous underscreening in the restricted primitive
  model}} (\bibinfo{year}{2022}),
  \urlprefix\url{https://arxiv.org/abs/2209.03486}.

\bibitem[{\citenamefont{Ciach}(2011)}]{ciach:11:2}
\bibinfo{author}{\bibfnamefont{A.}~\bibnamefont{Ciach}}, \bibinfo{journal}{{\it
  Mol. Phys}} \textbf{\bibinfo{volume}{109}}, \bibinfo{pages}{1101}
  (\bibinfo{year}{2011}).

\bibitem[{\citenamefont{Ciach}(2018)}]{ciach:18:0}
\bibinfo{author}{\bibfnamefont{A.}~\bibnamefont{Ciach}}, \bibinfo{journal}{{\it
  Soft Matter}} \textbf{\bibinfo{volume}{14}}, \bibinfo{pages}{5497}
  (\bibinfo{year}{2018}).

\bibitem[{\citenamefont{Ciach et~al.}(2003)\citenamefont{Ciach, G\'o\'zd\'z,
  and Evans}}]{ciach:03:1}
\bibinfo{author}{\bibfnamefont{A.}~\bibnamefont{Ciach}},
  \bibinfo{author}{\bibfnamefont{W.~T.} \bibnamefont{G\'o\'zd\'z}},
  \bibnamefont{and} \bibinfo{author}{\bibfnamefont{R.}~\bibnamefont{Evans}},
  \bibinfo{journal}{{\it J. Chem. Phys.}} \textbf{\bibinfo{volume}{118}},
  \bibinfo{pages}{3702} (\bibinfo{year}{2003}).

\bibitem[{\citenamefont{Outhwaite}(1975)}]{outhwaite:75:0}
\bibinfo{author}{\bibfnamefont{C.}~\bibnamefont{Outhwaite}},
  \emph{\bibinfo{title}{Statistical maechanics. A Specialist periodic Report}},
  vol.~\bibinfo{volume}{2} (\bibinfo{publisher}{The Chemical Society},
  \bibinfo{address}{London}, \bibinfo{year}{1975}), \bibinfo{note}{p.188}.

\bibitem[{\citenamefont{Ciach and Stell}(2005)}]{ciach:05:0}
\bibinfo{author}{\bibfnamefont{A.}~\bibnamefont{Ciach}} \bibnamefont{and}
  \bibinfo{author}{\bibfnamefont{G.}~\bibnamefont{Stell}},
  \bibinfo{journal}{{\it Int.J. Mod. Phys. B}} \textbf{\bibinfo{volume}{19}},
  \bibinfo{pages}{3309} (\bibinfo{year}{2005}).

\bibitem[{\citenamefont{Brazovskii}(1975)}]{brazovskii:75:0}
\bibinfo{author}{\bibfnamefont{S.~A.} \bibnamefont{Brazovskii}},
  \bibinfo{journal}{{\it Sov. Phys. JETP}} \textbf{\bibinfo{volume}{41}},
  \bibinfo{pages}{85} (\bibinfo{year}{1975}).

\bibitem[{\citenamefont{Mansoori et~al.}(1971)\citenamefont{Mansoori, Carnahan,
  Starling, and T.~W.~Leland}}]{Mansoori:71}
\bibinfo{author}{\bibfnamefont{G.}~\bibnamefont{Mansoori}},
  \bibinfo{author}{\bibfnamefont{N.~F.} \bibnamefont{Carnahan}},
  \bibinfo{author}{\bibfnamefont{K.~E.} \bibnamefont{Starling}},
  \bibnamefont{and}
  \bibinfo{author}{\bibfnamefont{J.}~\bibnamefont{T.~W.~Leland}},
  \bibinfo{journal}{{\it J. Chem. Phys.}} \textbf{\bibinfo{volume}{54}},
  \bibinfo{pages}{1523} (\bibinfo{year}{1971}).

\end{thebibliography}

\end{document}